\documentclass[12pt]{article}
\usepackage{graphicx}

\usepackage{times}

\newenvironment{sciabstract}{%
\begin{quote} \bf}
{\end{quote}}

\title{Absence of the $T^{2/3}$ specific heat anomaly in a $U(1)$ spin liquid with a large spinon Fermi surface: confinement of slave particles and non-locality of spinons}
\author{Tao Li$^{1,*}$\\
\\
\normalsize{$^{1}$Department of Physics, Renmin University of China,}\\
\normalsize{Beijing, 100872, P.R.China}\\
\\
\normalsize{$^{*}$E-mail:  litao\_phys@ruc.edu.cn.}
}

\date{}

\begin{document}

\baselineskip24pt

\maketitle

\begin{sciabstract}
Effective gauge theories based on slave particle construction are widely used to describe quantum number fractionalization in strongly correlated electron systems. However, even setting aside the intense debates on the confinement issue of the slave particles, there are still significant conflicts between theory and experiment. In particular, a $T^{2/3}$ specific heat anomaly has been predicted as the key signature of low-lying gauge fluctuation in a $U(1)$ spin liquid with a large spinon Fermi surface, which is however never observed. Here we show that such an anomaly is actually an artifact of a Gaussian approximation and is absent when the no double occupancy constraint on the slave particles is strictly enforced. We also show that projective construction based on slave particle representation provides a unified understanding on the mechanism of  spin fractionalization and the nonlocal nature of a physical spinon in one and two dimensional spin liquids.
\end{sciabstract}

\clearpage
\section*{Introduction}

Quantum spin liquids are exotic states of matter that can host fractionalized quasiparticles\cite{PALee1,Balents}. Quantum number fractionalization has been suggested to provide an exotic interpretation for the anomalous dynamics observed in many quantum magnets that are hard to explain within the traditional spin wave theory. It also offers a novel mechanism for the non-Fermi liquid behavior observed in the cuprate superconductors. Effective gauge theory based on slave particle construction is the most widely used theoretical tool to describe quantum number fractionalization in strongly correlated electron systems.

The $U(1)$ spin liquid with a large spinon Fermi surface is a particular example of systems showing quantum number fractionalization.  Such a state can be understood roughly as the descendant of a metallic state near a Mott transition, in which electron correlation has already opened a charge gap while leaving the electron Fermi surface intact. An insulator with a large Fermi surface is exotic in the sense that the gapless quasiparticles on the Fermi surface should carry only the spin but not the charge quantum number of an electron and is intrinsically fractionalized. Indeed, in organic Mott insulators with a triangular lattice, people do find evidence for the existence of such a quantum spin liquid near the Mott transition\cite{Kanoda1,Kanoda2,Matsuda,Zhou}. Magnetic susceptibility and specific heat measurement at low temperature on such systems exhibit typical behavior of a Fermi liquid metal with a finite density of state on the Fermi surface. Such a picture is also supported by theoretical studies. Variational studies find that when the multi-spin exchange is strong enough, as is expected near a Mott transition, a $U(1)$ spin liquid state with a large spinon Fermi surface is the best variational ground state of a quantum antiferromagnet defined on the triangular lattice\cite{Motrunich}. Effective field theory study based on slave particle construction also arrives at the same conclusion in the saddle point approximation\cite{SSLee1}. 

However, one encounters serious problems when trying to go beyond the saddle point approximation. The effective theory of the above $U(1)$ spin liquid has the form of a compact $U(1)$ gauge field coupled to Fermionic slave particles that form a large Fermi surface\cite{Zhou}. It is well known that in 2+1 dimension a pure compact $U(1)$ gauge field is always confining as a result of the proliferation of singular gauge field configuration called instanton\cite{Ployakov}. It has been strongly debated if the instanton effect can be suppressed by the dissipative coupling to a gapless Fermion system and if the gauge non-neutral slave particle can appear in physical spectrum\cite{PALee2,Millis,Polchinski,PALee3,Nayak,Wen1,Herbut1,Herbut2,Hermele,SSLee2,Sachdev1,Senthil1}. Even if the instanton effect can indeed be suppressed, there are still strong conflicts between theory and experiment. The non-compact $U(1)$ gauge field in the Gaussian effective theory, which has no intrinsic dynamics of its own, will acquire a relaxational dynamics with a dynamical exponent $z=3$ as a result of the dissipative coupling to the current of the gapless Fermionic slave particles\cite{Reizer,PALee2}. In two dimension(2D), such an ultra-slow dynamics in the gauge fluctuation will result in a $T^{2/3}$ anomaly in the low temperature specific heat\cite{Wong}. This smoking-gun signature of the Gaussian effective theory, however, has never been observed in any serious experimental investigation\cite{Zhou}. These unresolved issues cast serious doubt on our identification of the organic Mott insulators as $U(1)$ spin liquid materials\cite{Senthil2,Grover,Sachdev2,Yao,Xu}.  

We note, however, confinement of slave particles does not necessarily imply the instability of a $U(1)$ spin liquid and the forbiddance of spin fractionalization. For example, it is well known that in one dimension(1D), in which gauge non-neutral particles are always confined, fractionalized spin excitations can emerge as domain walls in the spin correlation pattern. Most theorists think this mechanism of spin fractionalization is fundamentally different from the mechanism by de-confinement of slave particles, since the slave particles are local objects, while the domain wall excitations are topological in nature\cite{Fradkin1,Fradkin2,Nayak}. However, one still cannot help wondering if there is any unrevealed connection between the slave particles and the physical spinons. After all, the two share the same Fermi surface in the $U(1)$ spin liquid state.

In this paper, we reinvestigate these issues by combing effective field theory analysis and variational construction. We show that the Gaussian approximation to the gauge fluctuation in the effective theory of a $U(1)$ spin liquid is invalid as singular gauge field configurations always proliferate. We find that the dissipative coupling between the transverse $U(1)$ gauge field and the current of the slave particles is prohibited when the time component of the $U(1)$ gauge field is exactly integrated out. We find further that the dynamics of the transverse gauge fluctuation in the $U(1)$ spin liquid is determined by its coupling to the scalar spin chirality, which features a large characteristic energy throughout the Brillouin zone. The $T^{2/3}$ specific heat anomaly predicted by the Gaussian effective theory is thus absent. We also show that the Gutzwiller projection will transform the slave particle into a genuine nonlocal object, as a physical spinon should be, thanks to the Friedel sum rule and Anderson's theorem of orthogonality catastrophe. This unifies our understanding of spin fractionalization in1D and 2D spin liquids.

\section*{Effective gauge theory of a $U(1)$ spin liquid and the failure of the Gaussian approximation}
We start from the standard $U(1)$ gauge field formulation of a quantum antiferromagnet. For illustrative purpose, we consider the spin-$\frac{1}{2}$ antiferromagnetic Heisenberg model on the triangular lattice, 
\begin{eqnarray}
H=2J\sum_{<i,j>} \vec{\mathrm{S}}_{i}\cdot\vec{\mathrm{S}}_{j}.\nonumber
\end{eqnarray} 
Here the sum $\sum_{<i,j>}$ is over nearest neighboring bonds. In real materials, additional terms are needed to stabilize the $U(1)$ spin liquid state. Such terms will not change the discussion that will follow and we will include them at a later time.
 
To introduce the gauge field formulation of the problem, we represent the spins in terms of the Fermionic slave particles as $\vec{\mathrm{S}}_{i}=\frac{1}{2}\sum_{\alpha,\beta}f^{\dagger}_{i,\alpha}\vec{\sigma}_{\alpha,\beta}f_{i,\beta}$. To preserve the spin algebra, the slave particle should satisfy the constraint of no double occupancy of the form $\sum_{\alpha} f^{\dagger}_{i,\alpha}f_{i,\alpha}=1$.  This representation has a built-in $U(1)$ gauge redundancy, since the spin operator is unaffected when we perform a $U(1)$ gauge transformation of the form $f_{i,\alpha}\rightarrow e^{i\phi_{i}}f_{i,\alpha}$, where $\phi_{i}$  is an arbitrary $U(1)$ phase. 

In terms of the slave particles, the Hamiltonian can be rewritten as $H=-J\sum_{<i,j>}\hat{\chi}_{i,j}^{\dagger}\hat{\chi}_{i,j}$, with $\hat{\chi}_{i,j}=\sum_{\alpha}f^{\dagger}_{i,\alpha}f_{j,\alpha}$.  After the standard Hubbard-Stratonovich transformation on $\hat{\chi}_{i,j}$,  and assuming a uniform saddle point value of $\chi$ for $|\chi_{i,j}|$, which is believed to be gapped, the partition function of the system can be written as
  \begin{eqnarray}
 Z= Z_{0} \int \prod_{i,\mu,\tau,\alpha} \it{D}f^{\dagger}_{i,\alpha}(\tau)\it{D}f_{i,\alpha}(\tau)\it{D}a^{\mathrm{0}}_{i}(\tau) \it{D}a^{\mu}_{i}(\tau) e^{-S},\nonumber
 \end{eqnarray}
 in which 
 \begin{eqnarray}
S=\int_{0}^{\beta} d\tau   [\sum_{i,j,\alpha}f^{\dagger}_{i,\alpha}(\tau)G^{-1}_{i,j}(\tau) f_{j,\alpha}(\tau)-i\sum_{i}a^{0}_{i}(\tau)] .\nonumber
 \end{eqnarray} 
Here $G^{-1}_{i,j}(\tau)=[\partial_{\tau}+ia^{0}_{i}(\tau)]\delta_{i,j}-J\chi e^{ia^{\mu}_{i}(\tau)}$ is the inverse propagator of the slave particles in the presence of the auxiliary field $a^{\mu}_{i}$ and $a^{0}_{i}$, which are to be interpreted as the spatial and temporal component of a compact $U(1)$ gauge field. We note that $a^{0}_{i}(\tau)$ is a Lagrange multiplier introduced to enforce the no double occupancy constraint.  
The above form involves integration over huge number of pure gauge degree of freedoms. We can fix the gauge for $a^{\mu}_{i}$ and rewrite the partition function as
 \begin{eqnarray}
 Z= Z'_{0} \int \prod_{i,\mathrm{x},\tau,\alpha} \it{D}f^{\dagger}_{i,\alpha}(\tau)\it{D}f_{i,\alpha}(\tau)\it{D}a^{\mathrm{0}}_{i}(\tau) \it{D}\mathrm{\Phi}_{\mathrm{x}}(\tau) e^{-S}.\nonumber
 \end{eqnarray}
(see Supplementary Material A for more details on the derivation). Here $\mathrm{\Phi}_{\mathrm{x}}$ is the gauge flux enclosed in a triangle centered at $\mathrm{x}$. It is related to the scalar spin chirality on the triangle by  $\sin \Phi_{\mathrm{x}} \propto  <\hat{C}_{\mathrm{x}}>=<\vec{\mathrm{S}}_{i}\cdot (\vec{\mathrm{S}}_{j} \times \vec{S}_{k})> $.

In the Gaussian approximation, we approximate $ia^{0}_{i}(\tau)=\lambda$. The fluctuation of $a^{0}_{i}(\tau)$ around $\lambda$ is argued to be screened by the density response of the Fermion system and is neglected at low energy\cite{PALee2}. We are then left with a free Fermion system coupled to the transverse $U(1)$ gauge field at low energy. When the Fermion degree of freedom is integrated out, the transverse gauge field will acquire a relaxational dynamics with a dynamical exponent $z=3$ at low energy as the result of its dissipative coupling to the spinon current. In 2D, such an ultra-slow dynamics would imply a $T^{2/3}$ anomaly in the low temperature specific heat.
   
However, the treatment of $a^{0}_{i}(\tau)$ outlined above is not justified from either a physical or a mathematical point of view. When the no double occupancy constraint is strictly enforced by the integration over $a^{0}_{i}(\tau)$, the spinon current should vanish identically. Thus the coupling between the spinon current and the transverse gauge field is unphysical. At the same time, the projection to the subspace of no double occupancy is achieved by $\it{destructive}$ interference between the contributions to $Z$ from different gauge paths $a^{0}_{i}(\tau)$. One thus should not expect any single gauge path to dominate the partition function. To illustrate this point, we have calculated the contributions to $Z$ from different gauge paths. We find such contributions are unbounded in magnitude and strongly fluctuating in phase for a general gauge path(see Supplementary Material B for a proof). Saddle point approximation on such unbounded contributions is thus meaningless.

\section*{The gauge dynamics of a $U(1)$ spin liquid with a large spinon Fermi surface}
Anticipating the inadequacy of the Gaussian approximation, we integrate out $a^{0}_{i}(\tau)$ exactly. This leaves us with an effective theory for the gauge flux $\Phi_{\mathrm{x}}$, which takes the form of $Z= \int \prod_{\mathrm{x},\tau} \it{D}\mathrm{\Phi}_{\mathrm{x}}(\tau) e^{-\tilde{S}[\mathrm{\Phi}]}$, in which
 \begin{eqnarray}
e^{-\tilde{S}[\Phi]}= Z'_{0} \int \prod_{i,\tau,\alpha} \it{D}f^{\dagger}_{i,\alpha}(\tau)\it{D}f_{i,\alpha}(\tau)\it{D}a^{\mathrm{0}}_{i}(\tau)e^{-S}.\nonumber
\end{eqnarray}
Thus the effective action of the transverse gauge field is determined by the response from a projected Fermion system. To make further progress, we apply the saddle point approximation to the $\it{physical}$ gauge flux $\Phi_{\mathrm{x}}(\tau)$ and assume $\Phi_{\mathrm{x}}(\tau)=0$ at the saddle point. This saddle point corresponds to the $U(1)$ spin liquid state with a large spinon Fermi surface.  To study the fluctuation effect around such a saddle point,
we expand $\tilde{S}[\mathrm{\Phi}]$ around $\mathrm{\Phi}_{\mathrm{x}}(\tau)=0$ to the second order. The expansion reads
\begin{eqnarray}
\tilde{S}[\mathrm{\Phi}]\simeq \tilde{S}[0]+\int d\tau d\tau' \sum_{\mathrm{x,x'}} \mathrm{\Phi}_{\mathrm{x}}(\tau) K_{\mathrm{x,x'}}(\tau,\tau') \mathrm{\Phi}_{\mathrm{x'}}(\tau').\nonumber
\end{eqnarray}
It can be shown that the linear coupling between the $U(1)$ gauge potential and the spinon current vanishes identically as a result of the no double occupancy constraint. To the lowest order in $\chi$, what survives the Gutzwiller projection is a linear coupling between the $U(1)$ gauge flux and the scalar spin chirality. We thus have
\begin{eqnarray}
K_{\mathrm{x,x'}}(\tau,\tau')\propto -<\mathrm{T}_{\tau}\hat{C}_{\mathrm{x}}(\tau)\hat{C}_{\mathrm{x'}}(\tau')>.\nonumber
\end{eqnarray}
(see Supplementary Material C for the details of the proof). This is drastically different from the situation in the Gaussian effective theory, in which the gauge dynamics is determined by the current response of a free Fermion system.

A computation of the full spectrum of $\hat{C}_{\mathrm{x}}$ for the projected Fermion system is difficult. However, the center of gravity of the spectrum can be obtained easily from a sum rule analysis and is given {\it exactly\/} by
\begin{equation}
E_{\mathrm{q}} = \frac{1}{2}\frac{<G|[[\hat{C}_{\mathrm{q}},H],\hat{C}_{-\mathrm{q}}^{\dagger}]|G>}{<G|\hat{C}_{\mathrm{q}}\hat{C}_{-\mathrm{q}}^{\dagger}|G>},
\end{equation}
in which $\hat{C}_{\mathrm{q}}=N^{-1}\sum e^{i\mathrm{q}\cdot \mathrm{x}}\hat{C}_{\mathrm{x}}$ is the density of scalar spin chirality at momentum $\mathrm{q}$, $|G>$ is the ground state of the system in the saddle point approximation, which is nothing but the Gutzwiller projected Fermi sea state. We note that with $E_{\mathrm{q}}$ we can already judge the validity of  the Gaussian effective theory,  which predicts that the characteristic energy for long wave length gauge fluctuation should vanish like $q^{3}$.  If the Gaussian theory is indeed valid, one should expect $E_{\mathrm{q}}$ to vanish in the same way. More generally, in the Gaussian effective theory $E_{\mathrm{q}}$ should always vanish in the $q\to 0$ limit as a result of the $U(1)$ gauge symmetry of the Gaussian effective action. 

To check on this point, we have calculated $E_{\mathrm{q}}$ for the projected Fermi sea state on the triangular lattice assuming the following Hamiltonian
\begin{eqnarray}
H=J_{2}\sum_{<i,j>} P_{ij}+J_{4}\sum_{[i,j,k,l]}(P_{ijkl}+P_{ilkj}).\nonumber
\end{eqnarray}
Here $P_{ij}=2\vec{\mathrm{S}}_{i}\cdot\vec{\mathrm{S}}_{j}+1/2$ is the Heisenberg exchange coupling.  $P_{ijkl}$ is the four spin ring exchange around a rhombi $[i,j,k,l]$. $\sum_{[i,j,k,l]}$ denotes the sum over all elementary rhombi of the triangular lattice. As found by Motrunich\cite{Motrunich}, when $J_{4}\geq 0.3J_{2}$ the projected Fermi sea state is the best variational state of the model. Here we set $J_{4}=0.3J_{2}$ .  

The result of $E_{\mathrm{q}}$ is shown in Fig.1. In stark contrast to the prediction of the Gaussian effective theory, $E_{\mathrm{q}}$ is found to be strongly gapped throughout the Brillouin zone. This result can be understood by an inspection of the structure factor of the scalar spin chirality, which is shown in Fig.2. One find that the correlation of $\hat{C}_{\mathrm{x}}$ in real space is extremely short-ranged and the corresponding structure factor is almost featureless around $\mathrm{q}=0$. We note that the short-ranged nature of the correlation in $\hat{C}_{\mathrm{x}}$ has also been mentioned by Motrunich\cite{Motrunich}. 

A nonzero $E_{\mathrm{q}}$ does not necessarily imply a gapped gauge fluctuation spectrum. In fact, as is detailed in Supplementary Material D, the scalar spin chirality can excite either one, two or at most three pairs of particle-hole excitations on the spinon Fermi sea, whose spectral weight vanish as $\omega$, $\omega^{3}$ and $\omega^{5}$ at low energy. However, such local excitations can contribute at most a $T^{2}$ correction to the specific heat at low temperature. The $T^{2/3}$ specific heat anomaly predicted by the Gaussian effective theory is absent. 

\section*{The mechanism of spin fractionalization in 1D and 2D spin liquids and the nonlocal nature of a spinon}
The above result implies that the Gaussian effective theory for the $U(1)$ spin liquid is invalid and the physical spinon can not be understood as a deconfined slave particle. A natural question is then how the two are related. After all, they share the same Fermi surface in this $U(1)$ spin liquid state. This question has been addressed by Mudry and Fradkin more than two decades ago\cite{Fradkin1,Fradkin2}. They argued that, at least in 1D, the two are fundamentally different objects, since the physical spinon is then a topological object that corresponds to an anti-phase domain wall in the spin correlation pattern, while the slave particle is a local object. Here we show that the Gutzwiller projection will transform the slave particle into a nonlocal object that corresponds just to such an anti-phase domain wall.

The Gutzwiller projected Fermi sea state, namely $|G>=\mathrm{P_{G}}\prod_{\mathrm{|k|<k_{F}}}f^{\dagger}_{\mathrm{k},\uparrow}f^{\dagger}_{\mathrm{k},\downarrow}|0>=\mathrm{P_{G}}|\mathrm{FS}>$, is known to be a very accurate description of the ground state of the spin-$\frac{1}{2}$ antiferromagnetic Heisenberg chain model\cite{Ogata}. In fact, one should not be surprised by such an exactness from the gauge field theory formulation, since the only gauge field component in the case, $a^{0}_{i}(\tau)$, has been exactly integrated out through Gutzwiller projection. On a 1D ring with $N=4l+2$ sites, the wave function of $|\mathrm{FS}>$ is given by 
\begin{eqnarray}
\psi_{\mathrm{FS}}(\{i_{m}\},\{j_{n}\})=\psi_{s}\prod_{m<m'}(Z_{i_{m}}-Z_{i_{m'}})\prod_{n<n'}(Z_{j_{n}}-Z_{j_{n'}}),\nonumber
\end{eqnarray}
in which $\{i_{m}\}$ and $\{j_{n}\}$ are the sets of coordinates for the up and the down spin electrons, $Z_{i_{m}}=\exp(\frac{i 2\pi i_{m}}{N})$ is the chord coordinate of a lattice sites on the ring\cite{Laughlin}, $\psi_{s}=(\prod_{m,n}Z^{*}_{i_{m}}Z^{*}_{j_{n}})^{l}$. For this wave function, it can be shown that the change in phase when we exchange a up spin electron at site $i$ and a down spin electron at site $j$ is given by $N_{c}\pi$, where $N_{c}$ is the total electron number between site $i$ and site $j$\cite{Yang1}. When $|\mathrm{FS}>$ is projected to the subspace of no double occupancy, this phase structure reproduces the Marshall sign rule structure of the antiferromagnetic Heisenberg chain\cite{Marshall,Weng}(see Supplementary Material E for more details).

We now excite a pair of spinons on the ground state. Since the ground state of the system is constructed by Gutzwiller projection of the mean field ground state, one would naturally expect that Gutzwiller projection of the mean field excited state to provide a reasonable description of the excited state. Such a logic has been followed successfully by many groups in the literature\cite{Yunoki1,Nave,Yang2,Li1,Ivanov}.  Following this logic, the variational state for a pair of spinons excited at site $i$ and $j$ should have the form of
\begin{equation}
|i,j>=\mathrm{P_{G}}f^{\dagger}_{i,\uparrow}f_{j,\downarrow}|\mathrm{FS}>.
\end{equation}
As a result of the Gutzwiller projection, the wave function of $|i,j>$ in the Fock basis is given by the amplitude in $|\mathrm{FS}>$ with site $i$ empty, site $j$ doubly occupied and all other sites singly occupied. In other words, a spinon acts effectively as an impurity that generates either one more or one less available state as compared to the singly occupied background. According to the phase structure we proved for $\psi_{\mathrm{FS}}$, spin exchange across site $i$ or site $j$(but not both) in the spin chain would pick up an additional phase shift of $\pi$. This $\pi$ phase shift corresponds just to an anti-phase domain wall in the spin chain.

To extend this reasoning to 2D, we note that the above $\pi$ phase shift can actually be understood as the manifestation of the Friedel sum rule in 1D\cite{Friedel}, which claims that with the appearance of each additional available Fermion state within an 1D region, the phase of scattering amplitude across the region will change by $\pi$. In 2D, the Friedel sum rule equates the scattering phase shift on the Fermi surface with $\pi$ times the number of additional Fermion states generated by the impurity potential below the Fermi energy.  Thus, each spinon will contribute a phase shift of $\pi$ on the spinon Fermi surface and exert a nonlocal influence on the surrounding spin state. More specifically, according to Anderson's orthogonality theorem\cite{Anderson1,Anderson2,NDD}, we expect the spin state surrounding a spinon to be orthogonal to the ground state in the thermodynamic limit. This can be checked by computing the overlap between the two states. Since spinon can only be excited in pairs, whose total contribution to the phase shift on the Fermi surface is zero, we expect the overlap to vanish only when the separation between the spinons is infinite.   

We find that such an overlap is given by $O(i,j)=G(i,j)/\sqrt{p_{0,2}p_{\uparrow,\downarrow}}$ (see Supplementary Material  G for the details of the derivation), in which $G(i,j)=\sum_{\mathrm{|k|<k_{F}}}e^{i\mathrm{k}\cdot(\mathrm{R}_{i}-\mathrm{R}_{j})}$ is the free Fermion correlator,  which decays in 2D as $|\mathrm{R}_{i}-\mathrm{R}_{j}|^{-2}$ at large distance. $p_{0,2}$ and $p_{\uparrow,\downarrow}$ are given by $p_{0,2}=<\mathrm{FS'}| \mathrm{P}^{i}_{0}\mathrm{P}^{j}_{2} |\mathrm{FS'}>/<\mathrm{FS}| \mathrm{P_{G}}|\mathrm{FS}>$ and $p_{\uparrow,\downarrow}=<\mathrm{FS'}| \mathrm{P}^{i}_{\uparrow}\mathrm{P}^{j}_{\downarrow} |\mathrm{FS'}>/<\mathrm{FS}| \mathrm{P_{G}}|\mathrm{FS}>$,
in which $\mathrm{P}^{i}_{0},\mathrm{P}^{i}_{2}, \mathrm{P}^{i}_{\uparrow}$ and $\mathrm{P}^{i}_{\downarrow}$ are the projection operators for the empty, doubly occupied, up spin and down spin state on  site $i$. $|\mathrm{FS'}>=\prod_{i'\neq i,j}\mathrm{P}^{i'}_{\mathrm{G}}|\mathrm{FS}>$ is a partially projected Fermi sea.  

Since the spin correlation approaches zero in $\mathrm{P_{G}}|\mathrm{FS}>$ in the large distance limit, $p_{\uparrow,\downarrow}$ should approach $1/4$ in the same limit. What is less obvious is the long range behavior of $p_{0,2}$. At the mean field level, one find $p_{0,2}=p_{\uparrow,\downarrow}=1/4+G(i,j)$ and both approach $1/4$ in the large distance limit. To go beyond the mean field treatment, we have calculated $p_{0,2}$ and $p_{\uparrow,\downarrow}$ by the variational Monte Carlo method. The result is shown in Fig.3. One finds both $p_{0,2}$ and $p_{\uparrow,\downarrow}$ approach a finite(but now different) value in the large distance limit. Thus the overlap we are seeking is proportional to $G(i,j)$ and will vanish as $|\mathrm{R}_{i}-\mathrm{R}_{j}|^{-2}$ in the large distance limit. This proves the claimed orthogonality catastrophe upon spinon excitation in the $U(1)$ spin liquid state. We note that according to our construction, $p_{0,2}$ can actually be interpreted as the probability to separate a pair of spinons to the distance $|\mathrm{R}_{i}-\mathrm{R}_{j}|$. A non-vanishing value of $p_{0,2}$ in the large distance limit is thus consistent with the existence of free spinons.

\section*{Conclusions and outlooks}
In conclusion, we have shown that the Gaussian approximation to the gauge fluctuation is in general invalid in effective gauge theories of spin liquids based on slave particle construction. In particular, we find that the $U(1)$ spin liquid state with a large spinon Fermi surface on the triangular lattice is robust and the fluctuation in the transverse gauge field on this state features a large characteristic energy throughout the Brillouin zone. The $T^{2/3}$ anomaly in the specific heat predicted by Gaussian effective theories simply does not exist.  We also find that projective construction based on the slave particle representation provides a unified understanding on the mechanism of spin fractionalization in 1D and 2D spin liquids and on the nonlocal nature of a physical spinon. 

The results presented in this paper are of general relevance since effective gauge theories based on slave particle construction are widely used in the study of strongly correlated electron systems. Our results show that the constraint on the slave particle is always essential and that the effective action of the emergent gauge field can be totally different from that derived in the Gaussian approximation. In this work, we have developed a systematic way to find the effective gauge action beyond the Gaussian approximation, which can be applied to check previous theoretical predictions made on the basis of the Gaussian approximation. Two problems are particularly interesting in this respect. The first problem is the origin of the strange metal behavior of the optimally doped cuprates\cite{PALee2}. The second problem is the nature of the quantum disordered phase evolved from a Neel ordered state\cite{Read}. In both problems, emergent $U(1)$ gauge field plays an important role.

\section*{Acknowledgments}
The author acknowledges the support from NSFC Grant No. 11674391, 973 Project No. 2016YFA0300504, and the research fund from Renmin University of China. 

\section*{Supplementary Materials}
Supplementary Texts\\
Figs. S1 to S7\\
References \textit{(7,10,31-35,38-42)}

\clearpage

\begin{figure}[h!]
\includegraphics[width=16cm,angle=0]{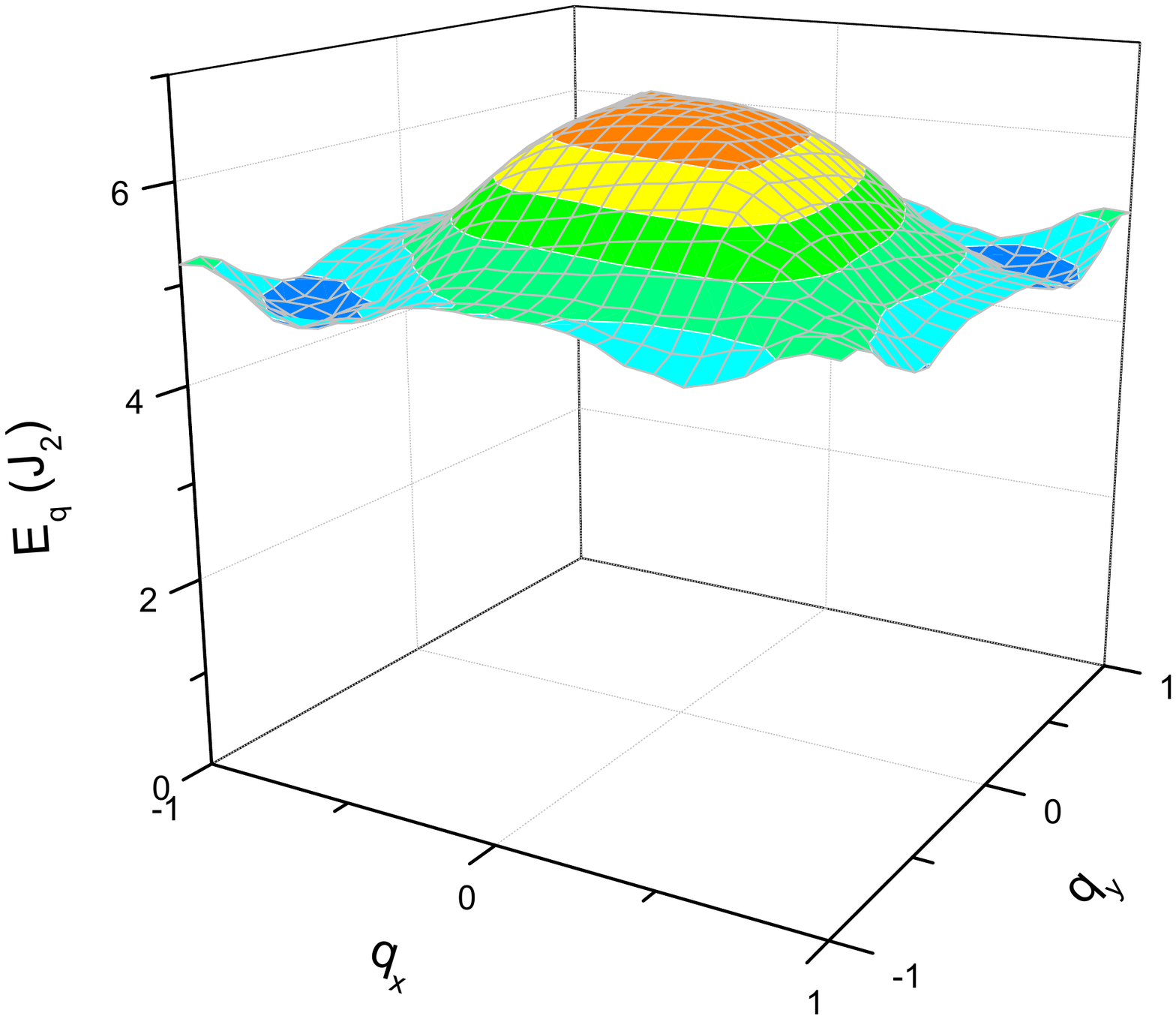}
\caption{The dispersion of $E_{\mathrm{q}}$ in the projected Fermi sea state on the triangular lattice. Shown here is the result for the acoustic mode in which the scalar spin chirality in the up and down triangles fluctuate in phase. We have adopted the convention $\vec{\mathrm{q}}=q_{x}\vec{\mathrm{G}}_{1}/2+q_{y}\vec{\mathrm{G}}_{2}/2$ for momentum, in which $\vec{\mathrm{G}}_{1,2}$ are the two reciprocal vectors of the triangular lattice.}
\end{figure}

\begin{figure}[h!]
\includegraphics[width=16cm,angle=0]{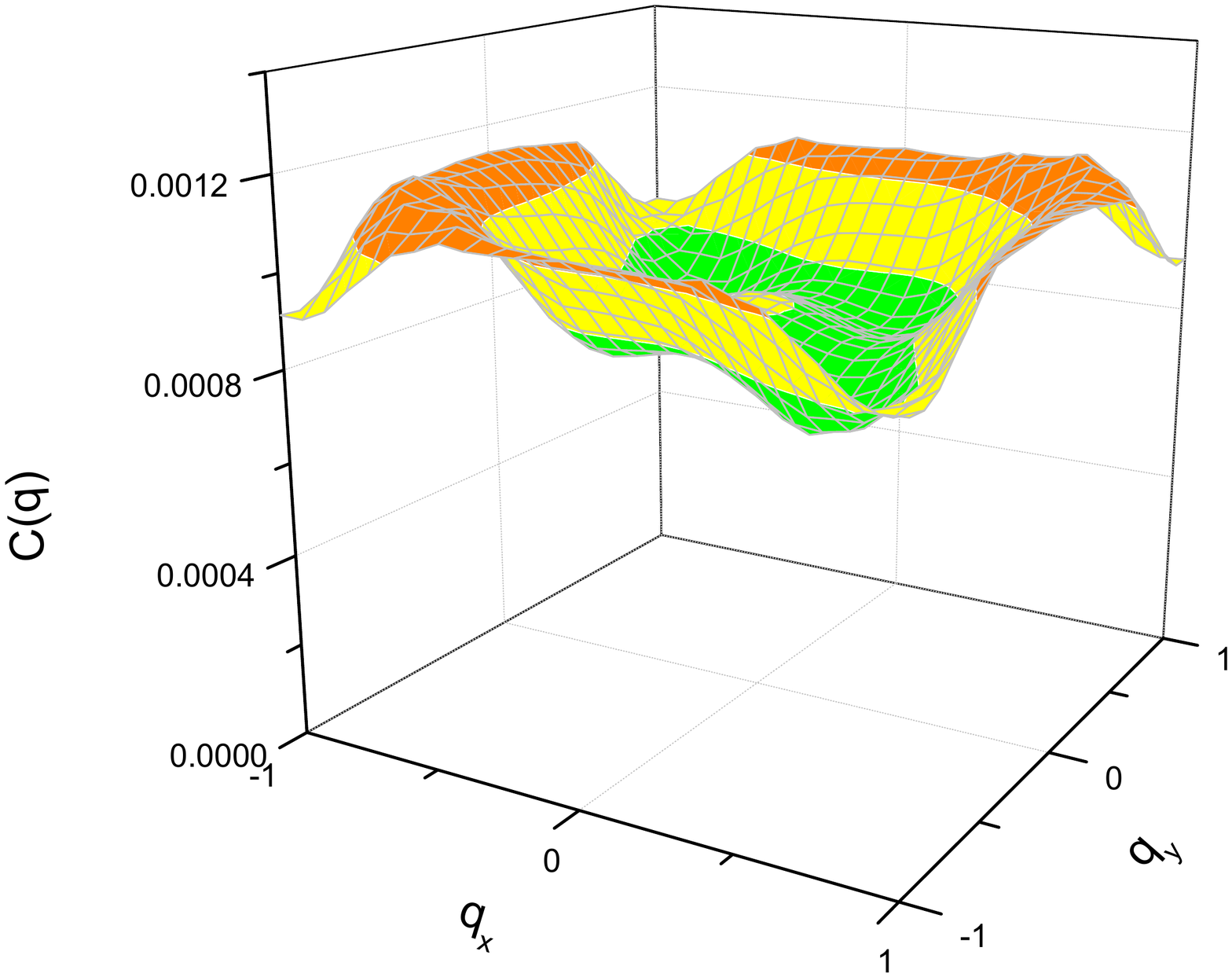}
\caption{The structure factor of the scalar spin chirality in the projected Fermi sea state on the triangular lattice. Shown here is the result for the acoustic mode.}
\end{figure}

\begin{figure}[h!]
\includegraphics[width=16cm,angle=0]{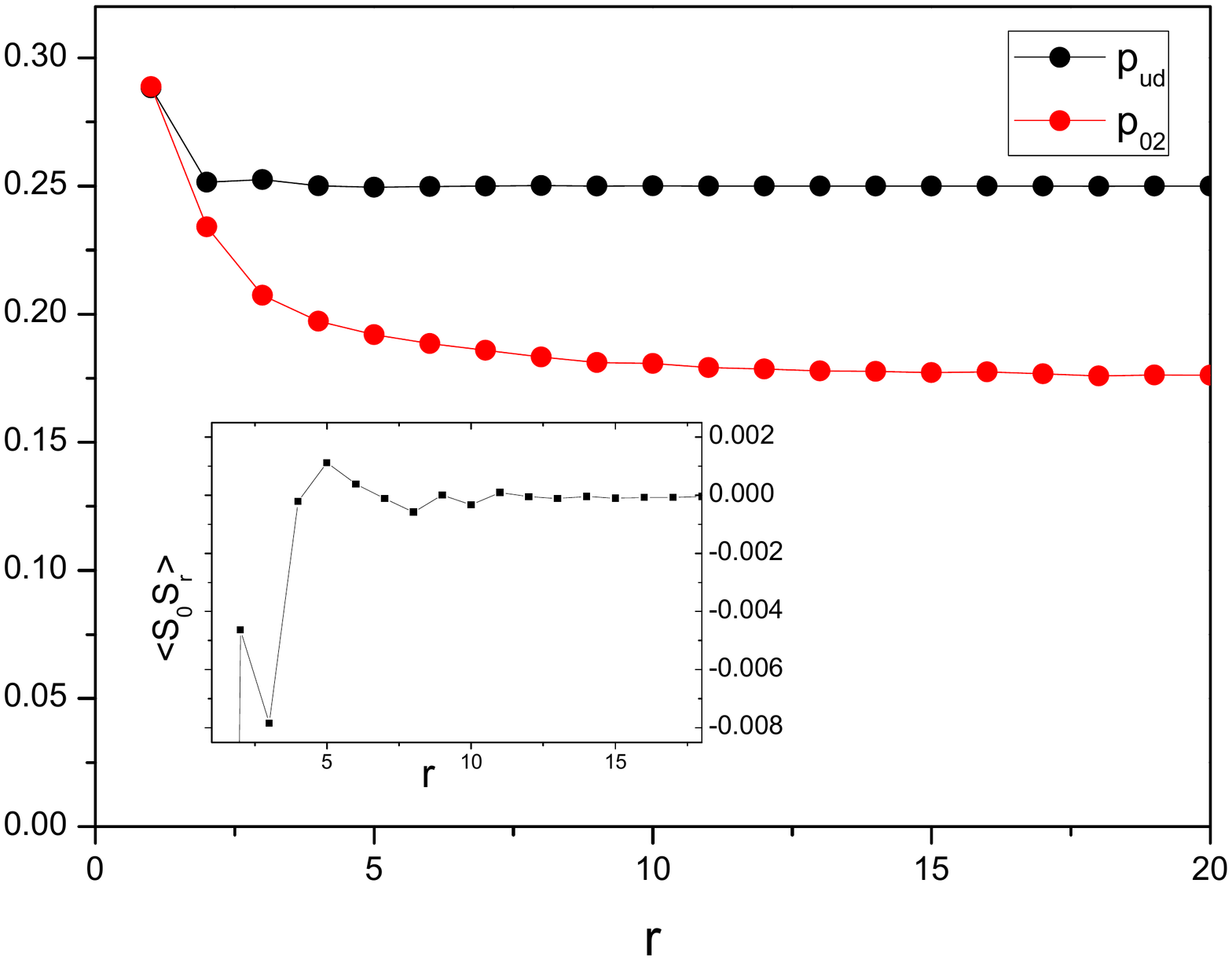}
\caption{The behavior of the function $p_{0,2}$ and $p_{\uparrow,\downarrow}$. The calculation is done on a $36\times36$ lattice. The inset shows the $2\mathrm{k_{F}}$ oscillation in $p_{\uparrow,\downarrow}$ in a magnified scale.}
\end{figure}

\clearpage
\centerline{\LARGE{Supplementary Materials for}}
\vspace{0.5cm}
\centerline{\Large{Absence of the $T^{2/3}$ specific heat anomaly in a $U(1)$ spin liquid with a large}}
\centerline{\Large{spinon Fermi surface:confinement of slave particles and non-locality of spinons}}
\centerline{Tao Li$^{1,*}$}
\centerline{\normalsize{$^{1}$Department of Physics, Renmin University of China,}}
\centerline{\normalsize{Beijing, 100872, P.R.China}}
\centerline{\normalsize{$^{*}$E-mail:  litao\_phys@ruc.edu.cn.}}

\renewcommand\thefigure{S\arabic{figure}}
\vspace{1cm}
\clearpage

\setcounter{equation}{0}

\section*{A. Derivation of the $U(1)$ effective gauge theory of the spin-$1/2$ quantum antiferromagnet on the triangular lattice}
We now derive an effective gauge field theory for the model introduced in the main text, which is given by({\it 7\/})
\begin{eqnarray}
H=J_{2}\sum_{<i,j>} P_{ij}+J_{4}\sum_{[i,j,k,l]}(P_{ijkl}+P_{ilkj}).\nonumber
\end{eqnarray}
Here $P_{ij}=2\vec{\mathrm{S}}_{i}\cdot\vec{\mathrm{S}}_{j}+1/2$ is the Heisenberg exchange coupling, $P_{ijkl}$ is the four spin ring exchange around a rhombi. $\sum_{[i,j,k,l]}$ denotes the sum over all elementary rhombi on the triangular lattice. In terms of the slave particles, the Hamiltonian can be written as({\it 7\/})
\begin{eqnarray}
&H&=J_{2}\sum_{<i,j>} (f^{\dagger}_{i,\alpha}f_{i,\beta})(f^{\dagger}_{j,\beta}f_{j,\alpha})\nonumber\\
&+&J_{4}\sum_{[i,j,k,l]}[(f^{\dagger}_{i,\alpha}f_{i,\beta})(f^{\dagger}_{j,\beta}f_{j,\gamma})(f^{\dagger}_{k,\gamma}f_{k,\delta})(f^{\dagger}_{l,\delta}f_{l,\alpha})+\mathrm{h.c.}]\nonumber.
\end{eqnarray}
Here and in the following, summation over repeated indices are assumed.  The slave particles should be subjected to the no double occupancy constraint to be a faithful representation of the spin algebra. 

In the coherent state path integral formulation, the partition function of the system can be written as({\it 10\/})
\begin{eqnarray}
 Z= \int \prod_{i,\tau,\alpha} \it{D}f^{\dagger}_{i,\alpha}(\tau)\it{D}f_{i,\alpha}(\tau)\it{D}a^{\mathrm{0}}_{i}(\tau)e^{-S},\nonumber
 \end{eqnarray}
 in which the action $S$ is given by
 \begin{eqnarray}
S=\int_{0}^{\beta} d\tau   [f^{\dagger}_{i,\alpha}(\tau)\partial_{\tau} f_{i,\alpha}(\tau)+H+ia^{0}_{i}(\tau)(f^{\dagger}_{i,\alpha}(\tau)f_{i,\alpha}(\tau)-1)]. \nonumber
 \end{eqnarray}  
 Here $a^{0}_{i}(\tau)$ is a Lagrange multiplier introduced to enforce the no double occupancy constraint. 
 
 We define the bond variable $\hat{\chi}_{i,j}=f^{\dagger}_{i,\alpha} f_{j,\alpha}$ and decouple the Heisenberg exchange term by the standard Hubbard-Stratonovich transformation on $\hat{\chi}_{i,j}$. The partition function after the transformation reads
 \begin{eqnarray}
 Z= \int \prod_{i,\tau,\alpha} \it{D}f^{\dagger}_{i,\alpha}(\tau)\it{D}f_{i,\alpha}(\tau)\it{D}a^{\mathrm{0}}_{i}(\tau)\it{D}\chi_{i,j}(\tau)e^{-S},\nonumber
 \end{eqnarray}
 in which the action $S$ is given by
 \begin{eqnarray}
S=\int_{0}^{\beta} d\tau   [f^{\dagger}_{i,\alpha}(\tau)G^{-1}_{i,j}(\tau) f_{j,\alpha}(\tau)+H_{4}-i\sum_{i}a^{0}_{i}(\tau)-J_{2}|\chi_{i.j}(\tau)|^{2}]. \nonumber
 \end{eqnarray}  
 Here $G^{-1}_{i,j}(\tau)=(\partial_{\tau}+ia^{0}_{i}(\tau))\delta_{i,j}-J_{2}\chi_{i,j}(\tau)$ is the inverse propagator of the slave particle in the presence of the auxiliary field $\chi_{i,j}(\tau)$ and $a^{0}_{i}(\tau)$, $H_{4}$ is the four spin exchange term left untouched. In the $U(1)$ spin liquid state, we can assume that the fluctuation in the amplitude of $\chi_{i,j}$ is gapped and can be neglected in low energy physics. It is then reasonable to assume $\chi_{i,j}\simeq \chi e^{ia^{\mu}_{i}}$, in which $\chi$ is a constant and $a^{\mu}_{i}$ is the phase of $\chi_{i,j}$. We thus have
 \begin{eqnarray}
 Z= Z_{0} \int \prod_{i,\mu,\tau,\alpha} \it{D}f^{\dagger}_{i,\alpha}(\tau)\it{D}f_{i,\alpha}(\tau)\it{D}a^{\mathrm{0}}_{i}(\tau) \it{D}a^{\mu}_{i}(\tau) e^{-S},\nonumber
 \end{eqnarray}
 in which
 \begin{eqnarray}
S=\int_{0}^{\beta} d\tau   [f^{\dagger}_{i,\alpha}(\tau)G^{-1}_{i,j}(\tau) f_{j,\alpha}(\tau)+H_{4}-i\sum_{i}a^{\mathrm{0}}_{i}(\tau)] \nonumber.
 \end{eqnarray} 
Here $G^{-1}_{i,j}(\tau)=(\partial_{\tau}+ia^{0}_{i}(\tau))\delta_{i,j}-J_{2}\chi e^{ia^{\mu}_{i}(\tau)}$.

The above form involves integration over huge number of pure gauge degree of freedoms. We can fix the gauge for the transverse gauge field and rewrite the partition function as
 \begin{eqnarray}
 Z= Z'_{0} \int \prod_{i,\mathrm{x},\tau,\alpha} \it{D}f^{\dagger}_{i,\alpha}(\tau)\it{D}f_{i,\alpha}(\tau)\it{D}a^{\mathrm{0}}_{i}(\tau) \it{D}\mathrm{\Phi}_{\mathrm{x}}(\tau) e^{-S}, \nonumber
 \end{eqnarray}
 in which $\Phi_{\mathrm{x}}$ is the $U(1)$ gauge flux enclosed in a triangle centered at $\mathrm{x}$. According to a well known identity({\it 31\/}), $\Phi_{\mathrm{x}}$ is related to the expectation value of the scalar spin chirality $\hat{C}_{\mathrm{x}}=\vec{\mathrm{S}}_{i}\cdot (\vec{\mathrm{S}}_{j} \times \vec{S}_{k})$ on triangle by  $\sin \Phi_{\mathrm{x}} \propto  <\vec{\mathrm{S}}_{i}\cdot (\vec{\mathrm{S}}_{j} \times \vec{S}_{k})>$, in which $i,j$ and $k$ are the three sites of the triangle. We note that we can also choose other gauge fixing conditions. For example, we can keep the integration over the gauge phase of the bond variable $\chi_{i,j}$, but require $a^{0}_{i}(\tau)$ to satisfy the condition $\partial_{\tau} a^{0}_{i}(\tau)=0$. However, we find that for the following discussion our choice of gauge condition is the most convenient one.

\section*{B. Failure of the saddle point approximation on $a^{0}_{i}(\tau)$ and the proliferation of singular gauge field configurations}

To begin with, we first demonstrate the failure of the saddle point approximation on $a^{0}_{i}(\tau)$ for a two-site toy model of the form $H=2\vec{\mathrm{S}}_{1}\cdot\vec{\mathrm{S}}_{2}$.  Following the general rule outlined above, one find the action of the system is given by
\begin{eqnarray}
S= \int_{0}^{\beta} &d\tau& [f^{\dagger}_{1,\alpha}(\tau)(\partial_{\tau}+ia^{0}_{1}(\tau))f_{1,\alpha}(\tau)\nonumber\\
 &+& f^{\dagger}_{2,\alpha}(\tau)(\partial_{\tau}+ia^{0}_{2}(\tau))f_{2,\alpha}(\tau)\nonumber\\
 &-&\chi (e^{ia^{1}(\tau)} f^{\dagger}_{1,\alpha}(\tau)f_{2,\alpha}(\tau)+\mathrm{h.c.})\nonumber\\
 &-&ia^{0}_{1}(\tau)-ia^{0}_{2}(\tau)].\nonumber
 \end{eqnarray}  
In the saddle point approximation, $ia^{0}_{i}(\tau)$ plays the role of chemical potential. As a result of the particle-hole symmetry of the action, the chemical potential is always zero at half filling. For our toy model, the spatial component of the gauge field, $a^{1}(\tau)$, can be gauged away(Note that this is also true for a 1D spin chain with open boundary, for which $a^{0}_{i}(\tau)$ is the only gauge field component that we need to consider). Thus the partition function of the toy model in the saddle point approximation is simply that of a two-level free Fermion system with eigenvalues $\chi$ and $-\chi$. 

Now we discretize the imaginary time into $N_{\tau}$ segments and calculate the contributions to the partition function from different gauge paths $a^{0}_{i}(\tau)$. To be more specific, we will calculate the contributions to $Z$ from gauge paths of the form $a^{0}_{i}(\tau)=z(i,\tau) N_{\tau}\pi/\beta$, in which $z(i,\tau)=0$ or $1$ is a random integer defined on the sites of the space-time lattice. The reason to choose such a special form can be understood as follows. As a result of the Pauli principle, the total number of Fermions on a given site can only be 0,1 and 2. Thus the projection into the singly occupied subspace can also be achieved by a discrete sum over all possible $z(i,\tau)$ configurations, rather than by an integration over the continuous Lagrange multiplier $a^{0}_{i}(\tau)$.  

The contribution of a given gauge path $a^{0}_{i}(\tau)$ to the partition function is given by({\it 32\/}) $C[a^{0}_{i}(\tau)]=\eta[\mathrm{Det} S]^{2}$, in which 
\begin{eqnarray}
S=\left(\begin{array}{cc}S_{1} &  S_{\chi} \\S_{\chi} & S_{2} \end{array}\right)\nonumber
\end{eqnarray}
is a $2N_{\tau}\times2N_{\tau}$ matrix, $\eta=\pm1$ is a sign determined by the parity of the sum $A=\sum_{i,\tau}z(i,\tau)$.  The submatrix $S_{i=1,2}$ and $S_{\chi}$ are given by
\begin{eqnarray}
S_{i}=\left(\begin{array}{ccccc}
1 & 0 & \cdots & 0 & a_{i,1} \\ 
-a_{i,2} & 1 & 0 & \ddots & 0 \\ 
0 & \ddots & \ddots & \ddots & \vdots\\ 
\vdots & \ddots & \ddots & \ddots & 0 \\ 
0 & \cdots & 0 & -a_{i,N_{\tau}} & 1 
\end{array}\right)\nonumber
\end{eqnarray} 
and
\begin{eqnarray}
S_{\chi}=\frac{-\beta\chi}{N_{\tau}}\left(\begin{array}{ccccc}
0 & 0 & \cdots & 0 & -1 \\ 
1 & 0 & 0 & \ddots & 0 \\ 
0 & \ddots & \ddots & \ddots & \vdots\\ 
\vdots & \ddots & \ddots & \ddots & 0 \\ 
0 & \cdots & 0 & 1 & 0 
\end{array}\right),\nonumber
\end{eqnarray} 
in which $a_{i,i_{\tau}}=1+iz(i,\tau)\pi$.

When $z(i,\tau)=0$, we should recover the contribution to the partition function from the saddle point, which is given by $C[z(i,\tau)=0]=4(1+\cosh(\beta\chi))^{2}$. For $N_{\tau}=200$, we find the truncation error in $C[z(i,\tau)]$ is about $5\times10^{-3}$ at $\beta\chi=1$.  For a random gauge path $z(i,\tau)$, we find the contribution to the partition function is strongly fluctuating in phase and unbounded in magnitude. In fact, we find that the amplitude of such contributions increases almost exponentially with the sum  $A=\sum_{i,\tau}z(i,\tau)$, as is illustrated in Fig. S1 for 1000 randomly chosen gauge paths. The maximum of $|C[z(i,\tau)]|$ is found to be achieved at $z(i,\tau)=1$, which is more than 400 orders of magnitude larger than the saddle point contribution for $N_{\tau}=200$. In fact, one can show that the contribution from this gauge path is given exactly by 
\begin{eqnarray}
C=(-1)^{A}(1+(1+\beta\chi/N_{\tau}+i\pi)^{N_{\tau}})^{2}\times(1+(1-\beta\chi/N_{\tau}+i\pi)^{N_{\tau}})^{2},\nonumber
\end{eqnarray}
which diverges as $(1+i\pi)^{4N_{\tau}}$ for large $N_{\tau}$. More generally, we note that in the $N_{\tau}\to \infty$ limit, the details in $S_{\chi}$ becomes immaterial to the value of the determinant $\mathrm{Det}S$, which can then be approximated by
\begin{eqnarray}
\mathrm{Det}S\simeq \prod_{i=1,2}[1+\prod_{i_{\tau}=1,N_{\tau}}(1+iz(i,\tau)\pi))].\nonumber
\end{eqnarray}
This explains the approximate exponential increase of $|\mathrm{Det}S|$ with $A$ shown in Fig. S1. 

The same reasoning can be easily extended to the case of a general lattice model. For example, the contribution from the gauge path $a^{0}_{i}(\tau)= N_{\tau}\phi/\beta$ to $Z$ is found to be given exactly by
\begin{eqnarray}
C=e^{iN_{s}N_{\tau}\phi}\prod_{\mathrm{k}}[1+(1+i\phi-\frac{\beta \epsilon_{\mathrm{k}}}{N_{\tau}})^{N_{\tau}}]^{2},\nonumber
\end{eqnarray}
in which $\epsilon_{\mathrm{k}}$ denotes the mean field eigenvalue of the lattice model, $N_{s}$ is the number of lattice sites. In the large $N_{\tau}$ limit, we find $C\simeq e^{iN_{s}N_{\tau}\phi}(1+i\phi)^{2N_{\tau}N_{s}}$. Such a contribution also diverges in the large $N_{\tau}$ limit. More generally, for an arbitrary gauge path $a^{0}_{i}(\tau)=N_{\tau}\phi_{i}(\tau)/\beta$, the details in the Hamiltonian is again immaterial if $\phi_{i}(\tau)$ remain finite in the $N_{\tau}\to \infty$ limit. We thus find
\begin{eqnarray}
C[a^{0}_{i}(\tau)]\simeq e^{i\sum_{i,i_{\tau}}\phi_{i}(\tau)}\prod_{i}[1+\prod_{i_{\tau}=1,N_{\tau}}(1+i\phi_{i}(\tau)))]^{2}.\nonumber
\end{eqnarray}
This is obviously unbounded in magnitude and strongly fluctuating in phase. Saddle point approximation on such contributions is meaningless.

In the $N_{\tau}\to \infty$ limit, a gauge path with $\phi_{i}(\tau)$ finite is singular. Such singular gauge field configurations are related(but not equivalent) to instantons of the $U(1)$ gauge field. For example, a gauge path of the form $a^{0}_{i}(\tau)=(2\pi N_{\tau}/\beta) \delta(\tau-\tau_{0}) \theta(y-y_{0})$ corresponds to a Dirac string of strength $2\pi$ running in the x direction, which can be understood as the remnant of a pair of oppositely charged instantons when they are annihilated after traversing the $x$-circumference of the system once.

\section*{C. The effective action of the transverse gauge field coupled to a projected Fermion system}
In the main text, we find that the effective action for the gauge flux $\Phi_{\mathrm{x}}(\tau)$ is determined by 
 \begin{eqnarray}
e^{-\tilde{S}[\Phi]}= Z'_{0} \int \prod_{i,\tau,\alpha} \it{D}f^{\dagger}_{i,\alpha}(\tau)\it{D}f_{i,\alpha}(\tau)\it{D}a^{\mathrm{0}}_{i}(\tau)e^{-S},\nonumber
\end{eqnarray}
in which
 \begin{eqnarray}
S=\int_{0}^{\beta} d\tau   [f^{\dagger}_{i,\alpha}(\tau)G^{-1}_{i,j}(\tau) f_{j,\alpha}(\tau)+H_{4}-i\sum_{i}a^{\mathrm{0}}_{i}(\tau)]. \nonumber
 \end{eqnarray} 
Here $G^{-1}_{i,j}(\tau)=(\partial_{\tau}+ia^{0}_{i}(\tau))\delta_{i,j}-J_{2}\chi e^{ia^{\mu}_{i}(\tau)}$.
Thus $\tilde{S}[\mathrm{\Phi}]$ is determined by the response of a projected Fermion system, rather than that of a free Fermion system. As we will show in the following, the responses of the two systems to the transverse gauge field are qualitatively different. In particular, while the transverse gauge field can couple directly to the Fermion current in the free Fermion system, it can only couple to the scalar spin chirality in the projected Fermion system to the lowest order of $\chi$.

For illustrative purpose, we neglect the four-spin ring exchange term $H_{4}$, which does not affect the discussion that will follow. We first rewrite the Fermion path integral representation of $e^{-\tilde{S}[\mathrm{\Phi}]}$ in the form of a trace over a series of Fock bases, which is given by
\begin{eqnarray}
e^{-\tilde{S}[\Phi]}= Z'_{0}\  \mathrm{Tr} \prod_{i_{\tau}=1}^{N_{\tau}}<\{n_{i_{\tau}+1}\}|\mathrm{P_{G}}e^{-\Delta\tau H^{\chi}_{i_{\tau}}}\mathrm{P_{G}}|\{n_{i_{\tau}}\}>.\nonumber
\end{eqnarray}
Here $|\{n_{i_{\tau}}\}>$ denotes a Fock basis at time $\tau=i_{\tau}\Delta\tau$, $\mathrm{Tr}$ indicates summation over all possible Fock bases $|\{n_{i_{\tau}}\}>$ that satisfy the condition $|\{n_{N_{\tau}}\}>=|\{n_{1}\}>$. The integration over the Lagrange multiplier $a^{0}_{i}(\tau)$ has been replaced by the Gutzwiller projection $\mathrm{P_{G}}$ on the Fock bases. $H^{\chi}_{i_{\tau}}$ is given by 
\begin{eqnarray}
H^{\chi}_{i_{\tau}}=-J_{2}\chi\sum_{<i,j>,\alpha}(e^{ia^{\mu}_{i}(\tau)}f^{\dagger}_{i,\alpha}f_{j,\alpha}+\mathrm{h.c.}).\nonumber
\end{eqnarray}

We now seek a quadratic approximation for $\tilde{S}[\mathrm{\Phi}]$ around the $U(1)$ spin liquid saddle point $\mathrm{\Phi}_{\mathrm{x}}(\tau)=0$, which takes the form of
\begin{eqnarray}
\tilde{S}[\mathrm{\Phi}]\simeq \tilde{S}[0]+\int d\tau d\tau' \sum_{\mathrm{x,x'}} \mathrm{\Phi}_{\mathrm{x}}(\tau) K_{\mathrm{x,x'}}(\tau,\tau') \mathrm{\Phi}_{\mathrm{x'}}(\tau').\nonumber
\end{eqnarray}
Denoting $Z[\mathrm{\Phi}]=e^{-\tilde{S}[\mathrm{\Phi}]}$, the kernel of the quadratic approximation for $\tilde{S}[\mathrm{\Phi}]$ is given by
\begin{eqnarray}
K_{\mathrm{x,x'}}(\tau,\tau')= - \frac{\delta^{2} \ln Z[\mathrm{\Phi}]}{\delta\mathrm{\Phi}_{\mathrm{x}}(\tau)\delta\mathrm{\Phi}_{\mathrm{x}'}(\tau')}. \nonumber
\end{eqnarray}
To find the kernel $K$, we expand $e^{-\Delta\tau H^{\chi}_{i_{\tau}}}$ in $\mathrm{\Phi}_{\mathrm{x}}(\tau)$. The lowest order term in $\chi$ in the expansion is given by 
\begin{eqnarray}
H_{1}=-J_{2}\Delta\tau\chi  \sum_{i,\mu}a^{\mu}_{i}(\tau)j^{\mu}_{i},\nonumber
\end{eqnarray}
in which $j^{\mu}_{i}=-i\sum_{\alpha}(f^{\dagger}_{i,\alpha}f_{j,\alpha}-\mathrm{h.c.})$
is the Fermion current. However, such a term does not survive the Gutzwiller projection $\mathrm{P_{G}}$. One find that when the Gutzwiller projection is taken into account, to the lowest order in $\chi$ the expansion of $\mathrm{P_{G}}e^{-\Delta\tau H^{\chi}_{i_{\tau}}}\mathrm{P_{G}}$ in $\mathrm{\Phi}_{\mathrm{x}}(\tau)$ is  given by
\begin{eqnarray}
H_{1}=-\frac{2(J_{2}\Delta\tau\chi)^{3}}{3} \sum_{\mathrm{x}}\mathrm{\Phi}_{\mathrm{x}}(\tau)\hat{C}_{\mathrm{x}}.\nonumber
\end{eqnarray}
Here $\hat{C}_{\mathrm{x}}=\vec{\mathrm{S}}_{i}\cdot(\vec{\mathrm{S}}_{j}\times\vec{\mathrm{S}}_{k})=(P_{ijk}-P_{ikj})/4i$ is the scalar spin chirality on the triangle centered at $\mathrm{x}$. $i,j$ and $k$ are the three sites of the triangle,
\begin{eqnarray}
P_{i,j,k}=\sum_{\alpha,\beta,\gamma}(f^{\dagger}_{i,\alpha}f_{j,\alpha})(f^{\dagger}_{j,\beta}f_{k,\beta})(f^{\dagger}_{k,\gamma}f_{i,\gamma})\nonumber
\end{eqnarray}
is the three-spin ring exchange operator on the triangle. Here we note that an effective theory for $\mathrm{\Phi}_{\mathrm{x}}(\tau)$ is meaningful only for energy smaller than the characteristic energy of the fluctuation in $|\chi_{i,j}|$, which is of the order of $J_{2}$, we should have $J_{2}\Delta\tau\geq 1$. The coupling constant between $\mathrm{\Phi}_{\mathrm{x}}(\tau)$ and the scalar spin chirality is thus of order one. Thus to the lowest order in $\chi$, we have
\begin{eqnarray}
K_{\mathrm{x,x'}}(\tau,\tau')\propto -<\mathrm{T}_{\tau}\hat{C}_{\mathrm{x}}(\tau)\hat{C}_{\mathrm{x'}}(\tau')>.\nonumber
\end{eqnarray}

\section*{D. The gauge dynamics on the Gutzwiller projected Fermi sea state on the triangular lattice}
Unlike the current fluctuation in a free Fermion system, the fluctuation in the scalar spin chirality has a non-vanishing characteristic energy in the $\mathrm{q}\to 0$ limit. To illustrate this point, we have calculated the spectral function of $\hat{C}_{\mathrm{x}}$ in the $U(1)$ spin liquid state at the mean field level. In general, the scalar spin chirality operator $\hat{C}_{\mathrm{x}}$ can excite at most three pairs of particle-hole pairs on the Fermi sea state. This can be seen more directly by rewriting $\hat{C}_{\mathrm{x}}$ as the sum of normal-ordered operators with respect to the Fermi sea state. The expansion is given by
\begin{eqnarray}
\hat{C}_{\mathrm{x}}=:\hat{C}^{(1)}_{\mathrm{x}}:+:\hat{C}^{(2)}_{\mathrm{x}}:+:\hat{C}^{(3)}_{\mathrm{x}}:\nonumber
\end{eqnarray}
in which
\begin{eqnarray}
:\hat{C}^{(1)}_{\mathrm{x}}:=\frac{3\chi^{2}}{16i}:(\hat{\chi}_{i,j}+\hat{\chi}_{j,k}+\hat{\chi}_{k,i}-\mathrm{h. c.}):\nonumber
\end{eqnarray}
is proportional to the sum of Fermion current around the triangle in the anti-clockwise manner. 
\begin{eqnarray}
:\hat{C}^{(2)}_{\mathrm{x}}:&=&\frac{\chi}{4i}:(\hat{\chi}_{i,j}\hat{\chi}_{k,i}+\hat{\chi}_{j,k}\hat{\chi}_{i,j}+\hat{\chi}_{k,i}\hat{\chi}_{j,k}-\mathrm{h. c.}):\nonumber\\
&-&\frac{\chi}{8i}:(\hat{\chi}_{i,i}\hat{\chi}_{j,k}+\hat{\chi}_{j,j}\hat{\chi}_{k,i}+\hat{\chi}_{k,k}\hat{\chi}_{i,j}-\mathrm{h. c.}):,\nonumber
\end{eqnarray}
in which $\hat{\chi}_{i,i}=\sum_{\alpha}f^{\dagger}_{i,\alpha}f_{i,\alpha}$ is the particle number operator on site $i$.
\begin{eqnarray}
:\hat{C}^{(3)}_{\mathrm{x}}:=\frac{1}{4i}:(\hat{\chi}_{i,j}\hat{\chi}_{j,k}\hat{\chi}_{k,i}-\mathrm{h. c.}):\nonumber.
\end{eqnarray}
These terms excite respectively one, two and three pairs of particle-hole pairs on the Fermi sea state.  
Simple phase space argument indicates that the spectral weight corresponding to $:\hat{C}^{(1)}_{\mathrm{x}}:$, $:\hat{C}^{(2)}_{\mathrm{x}}:$and $:\hat{C}^{(3)}_{\mathrm{x}}:$ should vanish as $\omega$, $\omega^{3}$ and $\omega^{5}$ at low energy. In particular, the spectral weight corresponding to $:\hat{C}^{(1)}_{\mathrm{x}}:$ should be proportional to $\omega/v_{F}q$ at low energy and should have a upper cutoff at $v_{F}q$ in the long wavelength limit as a result of the Pauli principle. Here $v_{F}$ is the Fermi velocity on the Fermi surface. On the other hand, the excitation corresponding to $:\hat{C}^{(2)}_{\mathrm{x}}:$ and $:\hat{C}^{(3)}_{\mathrm{x}}:$ do not suffer from so strong a phase space limitation. Their spectral weights can thus extend to large energy even at $q=0$ and should depend only weakly on $q$. 
These arguments are illustrated in Fig. S2, in which we plot the spectral weight and the corresponding real part of the response function for $:\hat{C}^{(1)}_{\mathrm{x}}:$, $:\hat{C}^{(2)}_{\mathrm{x}}:$ and $:\hat{C}^{(3)}_{\mathrm{x}}:$ separately. From the plot we see in the long wavelength limit the main spectral weight of $\hat{C}_{\mathrm{x}}$ comes from $:\hat{C}^{(2)}_{\mathrm{x}}:$ and $:\hat{C}^{(3)}_{\mathrm{x}}:$, both of which are characterized by large energy scale and are only weakly momentum dependent. As a result, the real part of the response function of $\hat{C}_{\mathrm{x}}$ is dominated by the contribution from $:\hat{C}^{(2)}_{\mathrm{x}}:$ and $:\hat{C}^{(3)}_{\mathrm{x}}:$ at low energy and is almost momentum and frequency independent.

There is one more detail on the excitation by $:\hat{C}^{(1)}_{\mathrm{x}}:$. On the triangular lattice, there are two inequivalent triangles in each unit cell, namely the up and the down triangle. We thus should consider both the in-phase(acoustic) and the out-of-phase(optical) fluctuation of $\hat{C}_{\mathrm{x}}$ on these triangles. We note that the excitation of one particle-hole pair in the acoustic channel is suppressed by an additional factor of $q^{2}$ in the long wavelength limit as compared to that in the optical channel, since the sum of $:\hat{C}^{(1)}_{\mathrm{x}}:$ over all triangles of the triangular lattice is identically zero.   

With these understandings in mind, we can write down the asymptotic form of the inverse gauge propagator $K(q,\omega)$ in the low energy regime as 
\begin{eqnarray}
K(\mathrm{q},\omega)\simeq K(\mathrm{q},0)+\frac{i\alpha(q)\omega}{v_{F}q}\nonumber
\end{eqnarray}
Here $K(\mathrm{q},0)$ is the response function of $\hat{C}_{\mathrm{x}}$ at zero frequency. According to the discussion above it should be a weakly $q$ dependent real number and can be treated as a constant in the low energy regime.  $\alpha(q)$ is a coupling constant. For the acoustic mode, $\alpha(q)\propto q^{2}$ in the $q\to 0$ limit. For the optical mode, $\alpha(q)$ should be approximately a constant in the $q\to 0$ limit. We thus expect the gauge fluctuation in the acoustic and optical channel to contribute a $T^{4}$ and $T^{2}$ correction to the low temperature specific heat, which are both dominated by the linear in T contribution from single spinon excitation at low temperature.
  
We now go beyond the mean field treatment and consider the fluctuation spectrum of $\hat{C}_{\mathrm{x}}$ on the Gutzwiller projected Fermi sea state. Since $\hat{C}_{\mathrm{x}}$ is a gauge invariant quantity(it conserves the Fermion number on a given site), it commute with the Gutzwiller projection operator, namely
\begin{eqnarray}
\hat{C}_{\mathrm{x}}\mathrm{P_{G}}|\mathrm{FS}>=\mathrm{P_{G}}\hat{C}_{\mathrm{x}}|\mathrm{FS}>.\nonumber
\end{eqnarray}
We thus have
\begin{eqnarray}
\hat{C}_{\mathrm{q}}\mathrm{P_{G}}|\mathrm{FS}>=\mathrm{P_{G}} :\hat{C}^{(1)}_{\mathrm{q}}:|\mathrm{FS}>+\mathrm{P_{G}}:\hat{C}^{(2)}_{\mathrm{q}}:|\mathrm{FS}>+\mathrm{P_{G}}:\hat{C}^{(3)}_{\mathrm{q}}:|\mathrm{FS}>.\nonumber
\end{eqnarray}
Therefore the excitation picture of $\hat{C}_{\mathrm{x}}$ on the Gutzwiller projected Fermi sea state is exactly the same as what we have described above  in the mean field treatment, albeit we should replace the mean field excited states with their Gutzwiller projected counterparts. Thus, the fluctuation spectrum of scalar spin chirality on the projected Fermion sea state should be qualitatively the same as the mean field prediction. We note that the mean field eigenstates will in general no longer be orthonormal after the Gutzwiller projection. However, the mean field energetics will be qualitatively preserved after the projection({\it 38-42\/}).

While a computation of the full spectrum of $\hat{C}_{\mathrm{x}}$ for the projected Fermion system is difficult, the center of gravity of the spectrum can be easily obtained. As we mentioned in the main text, the center of gravity of the fluctuation spectrum is given by
\begin{eqnarray}
E_{\mathrm{q}} =\frac{1}{2}\frac{<G|[[\hat{C}_{\mathrm{q}},H],\hat{C}_{-\mathrm{q}}^{\dagger}]|G>}{<G|\hat{C}_{\mathrm{q}}\hat{C}_{-\mathrm{q}}^{\dagger}|G>},
\end{eqnarray}
in which $\hat{C}_{\mathrm{q}}=N^{-1}\sum e^{i\mathrm{q}\cdot \mathrm{x}}\hat{C}_{\mathrm{x}}$ is the density of scalar spin chirality at momentum $\mathrm{q}$, $|G>$ is the ground state of the system in the saddle point approximation, which is nothing but the Gutzwiller projected Fermi sea state. The Hamiltonian we will use is the $J_{2}-J_{4}$ model of the form
\begin{eqnarray}
H=J_{2}\sum_{<i,j>} P_{ij}+J_{4}\sum_{[i,j,k,l]}(P_{ijkl}+P_{ilkj}).\nonumber
\end{eqnarray}
As found by Motrunich({\it 7\/}), when $J_{4}\geq 0.3J_{2}$, the projected Fermi sea state is the best variational state of the model. In our calculation we set $J_{4}=0.3J_{2}$. 

When expanded in real space, both the numerator and the denominator in Eq.(1) are sum of expectation values of local operators. For example, a general term in the numerator is proportional to 
$<G|[[P_{i,j,k},P_{l,m}],P_{i',j'k'}]|G>$ or $<G|[[P_{i,j,k},P_{l,m,n,r}],P_{i',j',k'}]|G>$, while a general term in the denominator is given by $<G|P_{i,j,k}P_{i',j'k'}|G>$. Such expectation values can be easily calculated by the variational Monte Carlo method. In our calculation, we have used a $24\times24$ lattice with periodic-antiperiodic boundary condition for the slave particles. We have used $1.28\times 10^{7}$ statistically independent samples to calculate the double commutator and the structure factor in Eq.(1). Each sample is drawn after 1000 local updates. The statistical error of the data presented in our figures are already smaller than the symbol size. 

In the main text, we have presented the results for the acoustic gauge mode. For completeness, here we present the results for the optical gauge mode. In Fig. S3 and S4, we plot the center of gravity of the spectrum and the structure factor of the optical gauge mode. Except for the small spike in $E_{\mathrm{q}}$ at $\mathrm{q}=0$, the optical gauge mode is found to behave in a similar way as the acoustic gauge mode. Such a spike is caused by a related dip in the structure factor around the $\Gamma$ point and can be understood as the consequence of the Pauli principle on the one particle-hole excitation. In the acoustic channel, the one particle-hole continuum is suppressed by an additional factor of $q^{2}$ in the long wave length limit, making its momentum dependence not as obvious in the structure factor.  

We have also made a finite size scaling analysis of $E_{\mathrm{q}}$ for the acoustic gauge mode at $\mathrm{q}=0$. As shown in Fig. S5, $E_{\mathrm{q=0}}$ is almost independent of the lattice size when $L\geq 6$.

\section*{E. The wave function and its sign structure of the one dimensional projected Fermi sea state at half filling}
The content of this subsection is essentially reproduced from an earlier work of us({\it 35\/}). Let us consider the spin-$1/2$ antiferromagnetic Heisenberg chain with the Hamiltonian $H=J\sum_{i}\vec{\mathrm{S}}_{i}\cdot\vec{\mathrm{S}}_{i+1}$. It is well known that the 1D Gutzwiller projected Fermi sea state of the form $\mathrm{P_{G}}|\mathrm{FS}>=\mathrm{P_{G}}\prod_{\mathrm{|k|<k_{F}}}f^{\dagger}_{\mathrm{k},\uparrow}f^{\dagger}_{\mathrm{k},\downarrow}|0>$ is an extremely accurate variational state of this model({\it 33\/}). For example, the relative error in the ground state energy calculated from  $\mathrm{P_{G}}|\mathrm{FS}>$ is smaller than $0.2\%$.  In fact, one should not be surprised by such an exactness from the gauge field theory formulation presented above, since the only gauge field component in the case, $a^{0}_{i}(\tau)$, has been exactly integrated out through Gutzwiller projection(we note that the fluctuation in the amplitude of the bond variable, $|\chi_{i,j}|$, which is believed to be unimportant for long wavelength physics, is still only treated at the saddle point level). 

For convenience, we consider the state on a finite ring with $N=4l+2$ sites and with periodic boundary condition. The boundary condition is so chosen to guarantee a closed shell structure at half filling. In the Fock basis, the wave function of the half-filled Fermi sea state is given by({\it 34\/})
 \begin{eqnarray}
\psi(\{i_{m}\},\{j_{n}\})=\psi_{s}\prod_{m<m'}(Z_{i_{m}}-Z_{i_{m'}})\prod_{n<n'}(Z_{j_{n}}-Z_{j_{n'}})\nonumber
\end{eqnarray}
in which $\{i_{m}\}$ and $\{j_{n}\}$ are the sets of coordinates for the up and the down spin electrons, $Z_{i_{m}}=e^{i2\pi i_{m}/N}$ is the chord coordinate on the ring,  $\psi_{s}$ is a symmetric function given by $\psi_{s}=(\prod_{m,n}Z^{*}_{i_{m}}Z^{*}_{j_{n}})^{l}$. In the projected Fermi sea state, all sites should be occupied by one and only one electron of either spin.

Without loss of generality, let us exchange a up spin electron at site $i_{1}$ with a down spin electron at site $j_{1}$. The change in the phase of the wave function is given by
\begin{eqnarray}
    \Delta\Phi=\arg(\prod_{\alpha>1}\frac{Z_{i_{\alpha}}-Z_{j_{1}}}{Z_{i_{\alpha}}-Z_{i_{1}}}\prod_{l>1}
    \frac{Z_{j_{l}}-Z_{i_{1}}}{Z_{j_{l}}-Z_{j_{1}}}).\nonumber
\end{eqnarray}
Since $|Z_{i_{m}}|=1$, the chord coordinates are complex numbers living on a unit circle.
Then $\theta^{\alpha}_{i_{1},j_{1}}=\arg((Z_{i_{\alpha}}-Z_{j_{1}})/(Z_{i_{\alpha}}-Z_{i_{1}}))$ is
nothing but the angle in the segment $Z_{i_{1}}-Z_{j_{1}}$ in the
unit circle(see Fig. S6 for an illustration). Noting the fact that in a circle the angles in the same
segment equal one another and the sum of the opposite angles of
quadrilaterals equals $\pi$, one easily find that
$\Delta\Phi=N_{c}\pi$, in which $N_{c}$ denotes the number of
electrons between site $i_{1}$ and site $j_{1}$. Taking into account
the sign due to Fermion exchange, one find the change in the phase of the wave function is in
accordance with the Marshall sign rule, which claims that the phase of the wave function should change by $\pi$ if we exchange two spins in different sublattices.

Now suppose we introduce a pair of spinons at site $i$ and site $j$. Following the logic we have mentioned in the main text, the variational state in this case should have the form of
\begin{eqnarray}
|i,j>=\mathrm{P_{G}}f^{\dagger}_{i,\uparrow}f_{j,\downarrow}|\mathrm{FS}>\nonumber.
\end{eqnarray}
To be consistent with the no double occupancy constraint, site $i$ should be empty in the Fermi sea state before the action 
of $f^{\dagger}_{i,\uparrow}$. For the same reason, site $j$ should be doubly occupied in the Fermi sea state before the action 
of $f_{j,\downarrow}$. All other sites should be singly occupied as usual. Thus a spinon in between will change $N_{c}$ by $1$(or $-1$) between any two sites. As a result, an additional phase shift of $\pi$ will be picked up when we exchange two spins across a spinon. This $\pi$ phase shift is responsible for the topological nature of a spinon as an anti-phase domain wall in the antiferromagnetic Heisenberg chain.

\section*{F. The demonstration of orthogonality catastrophe in the projected Fermi sea state upon the excitation of a pair of spinons}

According to our construction scheme, the wave function for the state with a pair of spinons excited at site $i$ and site $j$ is given by the amplitude in $|\mathrm{FS}>$ with site $i$ empty, site $j$ doubly occupied and all other sites singly occupied. The existence of a spinon thus acts effectively as an impurity that generates either one more or one less available state in the otherwise singly occupied background. According to the Friedel sum rule, a spinon will thus exert a nonlocal influence on the surrounding spin state. In particular, a quasiparticle living on the spinon Fermi surface will acquire a phase shift of $\pi$ in the presence of a spinon.  Such a nonzero phase shift on the Fermi surface will result in Anderson's orthogonality catastrophe and we thus expect the spin state surrounding a spinon to be orthogonal to the ground state in the same region in the thermodynamic limit. However, since spinon excitation can only be excited in pairs, whose total contribution to the phase shift on the spinon Fermi surface is zero, we expect the overlap between the spin state surrounding the spinon pair and the ground state in the same region to approach zero only when the separation between the pair of spinons is infinity. This is what we call orthogonality catastrophe upon the excitation of a pair of spinons.

Now we calculate such an overlap. We first rewrite the state with a pair of spinons excited at site $i$ and $j$ more explicitly as 
\begin{eqnarray}
|i,j>&=&f^{\dagger}_{i,\uparrow}f_{j,\downarrow}\mathrm{P}^{i}_{0}\mathrm{P}^{j}_{2}\prod_{i'\neq i,j}\mathrm{P}^{i'}_{\mathrm{G}}|\mathrm{FS}>\nonumber\\
&=&f^{\dagger}_{i,\uparrow}f_{j,\downarrow}\mathrm{P}^{i}_{0}\mathrm{P}^{j}_{2}|\mathrm{FS'}>,\nonumber
\end{eqnarray}
in which $\mathrm{P}^{i}_{0}$, $\mathrm{P}^{i}_{2}$,$\mathrm{P}^{i}_{\uparrow}$, and $\mathrm{P}^{i}_{\downarrow}$ are the projection operators into the subspace of the empty, doubly occupied, up spin and down spin state on site $i$. $\mathrm{P}^{i}_{\mathrm{G}}$ is the Gutzwiller projection operator on site $i$. As a result of the conservation of total $S^{z}$, there are only two components of the ground state that can contribute to the overlap with $|i,j>$. They are given by $|\uparrow,\downarrow>=\mathrm{P}^{i}_{\uparrow}\mathrm{P}^{j}_{\downarrow}|\mathrm{FS'}>$ and  $|\downarrow,\uparrow>=\mathrm{P}^{i}_{\downarrow}\mathrm{P}^{j}_{\uparrow}|\mathrm{FS'}>$. Using inversion symmetry of the system, it is easy to show that these two components generate the same spin state in the region surrounding the spinon pair. Thus in the following we only consider the first component.

Following these reasonings, one find the overlap can be expressed as
\begin{eqnarray}
O(i,j)&=&\frac{<\uparrow,\downarrow|f_{j,\uparrow}f^{\dagger}_{j,\downarrow}|i,j>}{\sqrt{<\uparrow,\downarrow|\uparrow,\downarrow>}\sqrt{<i,j|i,j>}}\nonumber\\
&=&\frac{<\mathrm{FS'}| \mathrm{P}^{i}_{\uparrow}\mathrm{P}^{j}_{\downarrow}f^{\dagger}_{i,\uparrow}f_{j,\uparrow} \mathrm{P}^{i}_{0}\mathrm{P}^{j}_{2}|\mathrm{FS'}>}{\sqrt{<\mathrm{FS'}| \mathrm{P}^{i}_{\uparrow}\mathrm{P}^{j}_{\downarrow} |\mathrm{FS'}>}\sqrt{<\mathrm{FS'}| \mathrm{P}^{i}_{0}\mathrm{P}^{j}_{2} |\mathrm{FS'}>}}.\nonumber
\end{eqnarray}   
Using the identity $f^{\dagger}_{i,\uparrow}\mathrm{P}^{i}_{0}=\mathrm{P_{G}}^{i}f^{\dagger}_{i,\uparrow}$ and $f_{j,\uparrow}\mathrm{P}^{j}_{2}=\mathrm{P_{G}}^{i}f_{j,\uparrow}$, and the conservation of total $S^{z}$, the numerator can be simplified to $<\mathrm{FS}| \mathrm{P_{G}}f^{\dagger}_{i,\uparrow}f_{j,\uparrow} |\mathrm{FS}>$. Using the translational symmetry of the system it reduces further to $G(i,j)\times<\mathrm{FS}| \mathrm{P_{G}} |\mathrm{FS}>$, in which $G(i,j)=\sum_{\mathrm{|k|<k_{F}}}e^{i\mathrm{k}\cdot(\mathrm{R}_{i}-\mathrm{R}_{j})}$ is the correlator of the free Fermion. Thus the overlap we are seeking can be expressed as $O(i,j)=G(i,j)/\sqrt{p_{0,2}p_{\uparrow,\downarrow}}$, in which 
\begin{eqnarray}
p_{0,2}=\frac{<\mathrm{FS'}| \mathrm{P}^{i}_{0}\mathrm{P}^{j}_{2} |\mathrm{FS'}>}{<\mathrm{FS}| \mathrm{P_{G}}|\mathrm{FS}>}\nonumber\\
p_{\uparrow,\downarrow}=\frac{<\mathrm{FS'}| \mathrm{P}^{i}_{\uparrow}\mathrm{P}^{j}_{\downarrow} |\mathrm{FS'}>}{<\mathrm{FS}| \mathrm{P_{G}}|\mathrm{FS}>}.\nonumber
\end{eqnarray}

In the large distance limit, the spin correlation approaches zero in $\mathrm{P_{G}}|\mathrm{FS}>$. Thus $p_{\uparrow,\downarrow}$ should approach $1/4$ in the same limit. What is less obvious is the long range behavior of $p_{0,2}$. At the mean field level(with $|\mathrm{FS'}>$ approximated by $|\mathrm{FS}>$),  it is easy to show that $p_{0,2}=p_{\uparrow,\downarrow}=1/4+G(i,j)$
and both approach $1/4$ in the large distance limit. To go beyond the mean field treatment, we have computed $p_{0,2}$ and $p_{\uparrow,\downarrow}$ with the variational Monte Carlo method. It is found that the equality $p_{0,2}=p_{\uparrow,\downarrow}$ no longer hold. However, it is found that $p_{0,2}$ still approaches a nonzero value in the large distance limit(see Fig. 3 in the main text).  
Thus the overlap we are seeking is proportional to $G(i,j)$ and will vanish as $|\mathrm{R}_{i}-\mathrm{R}_{j}|^{-2}$ in the large distance limit. This proves the claimed orthogonality catastrophe upon spinon excitation in the $U(1)$ spin liquid state. 

We note according to our construction, $p_{0,2}$ can actually be interpreted as the probability to separate a pair of spinons to the distance $|\mathrm{R}_{i}-\mathrm{R}_{j}|$. A non-vanishing value of $p_{0,2}$ in the large distance limit is thus consistent with the existence of free spinon.

\clearpage

\setcounter{figure}{0}
\begin{figure}[h!]
\includegraphics[width=16cm,angle=0]{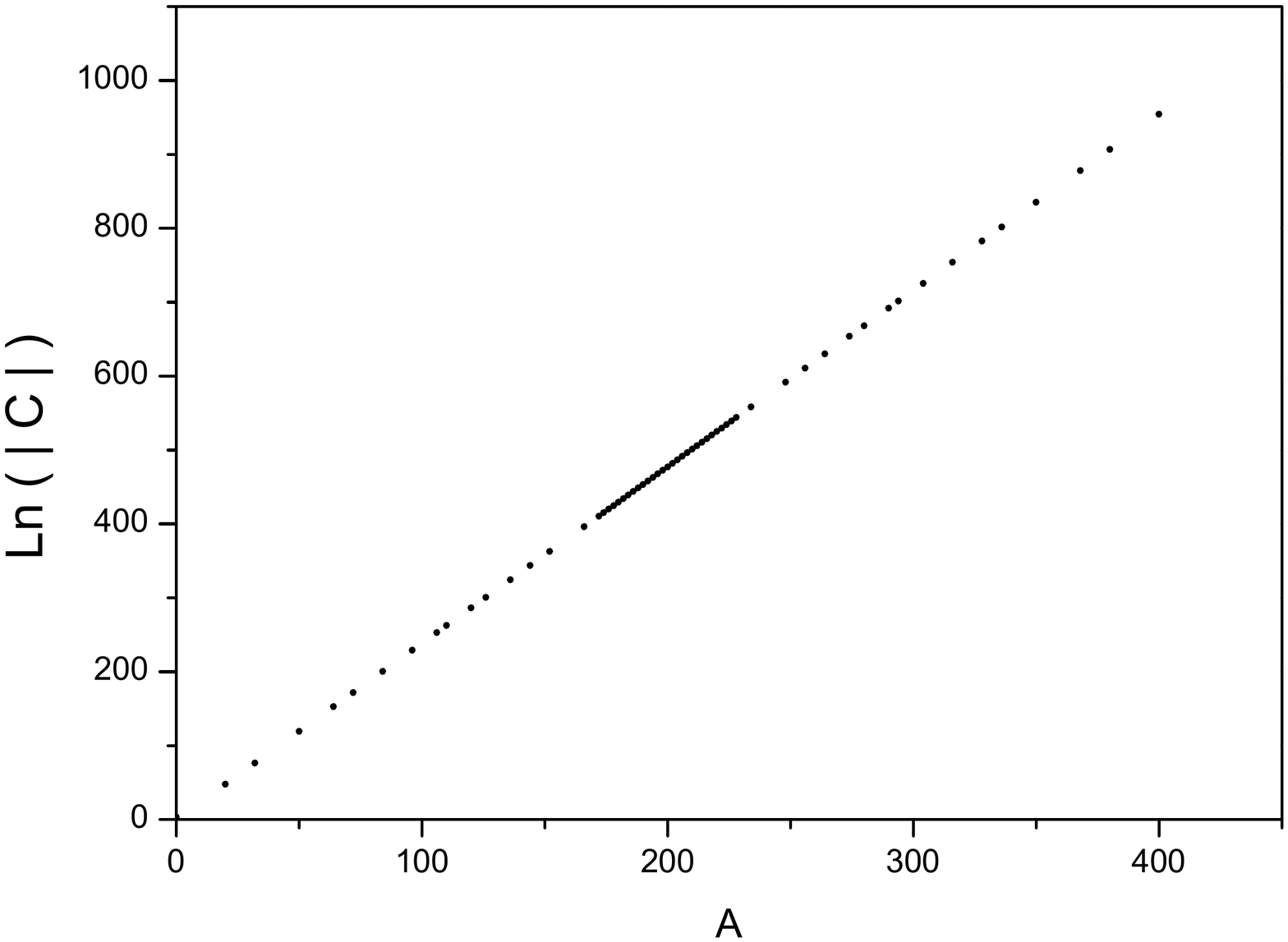}
\caption{The dependence of $\mathrm{Ln}(|C|)$ on the sum $A=\sum_{i,i_{\tau}}z(i,\tau)$ for 1000 randomly chosen gauge paths. Here we set $N_{\tau}=200$, $\beta=1$, $\chi=1$.}
\end{figure}

\begin{figure}[h!]
\includegraphics[width=16cm,angle=0]{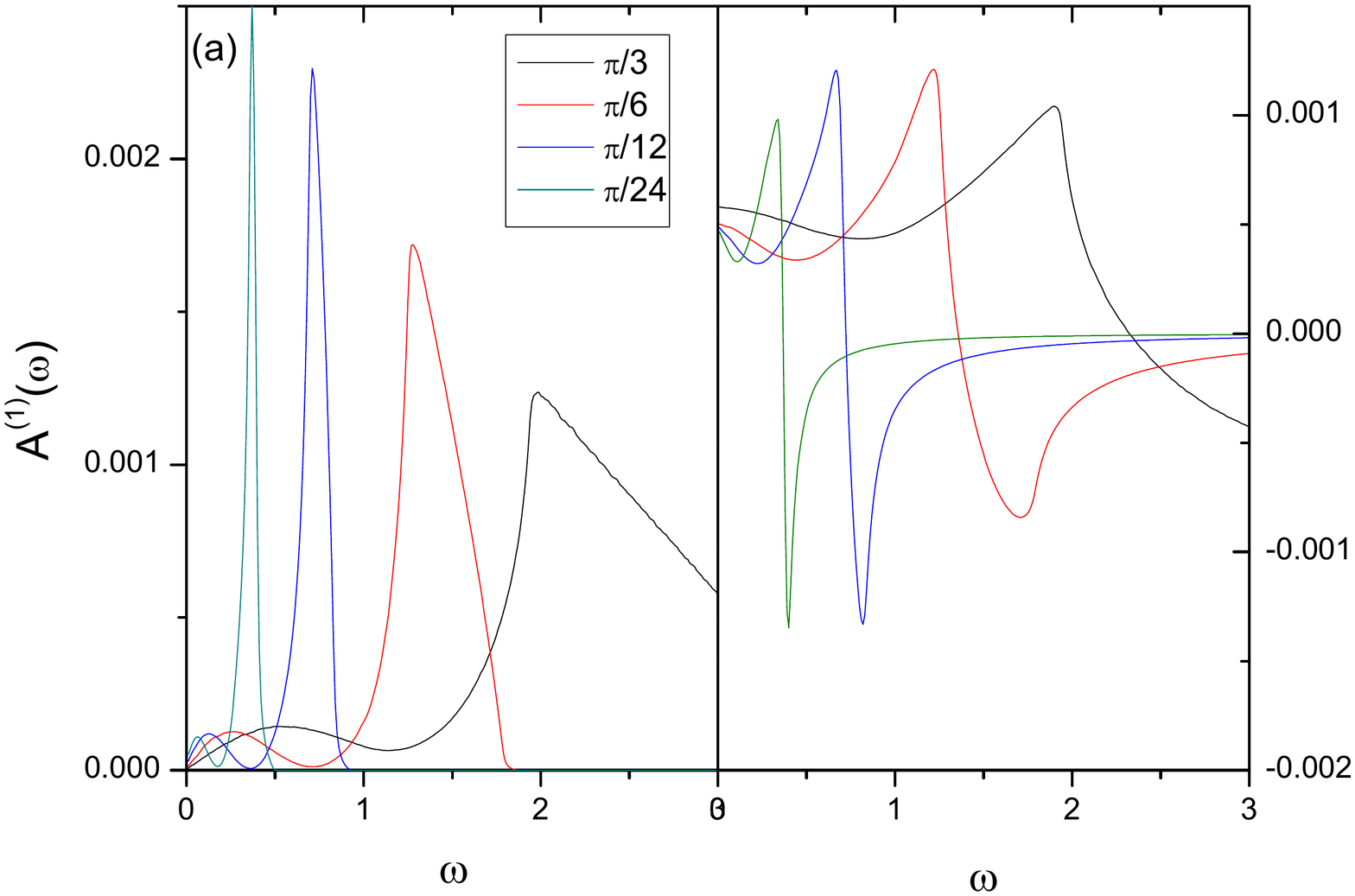}
\end{figure}
\begin{figure}[h!]
\includegraphics[width=16cm,angle=0]{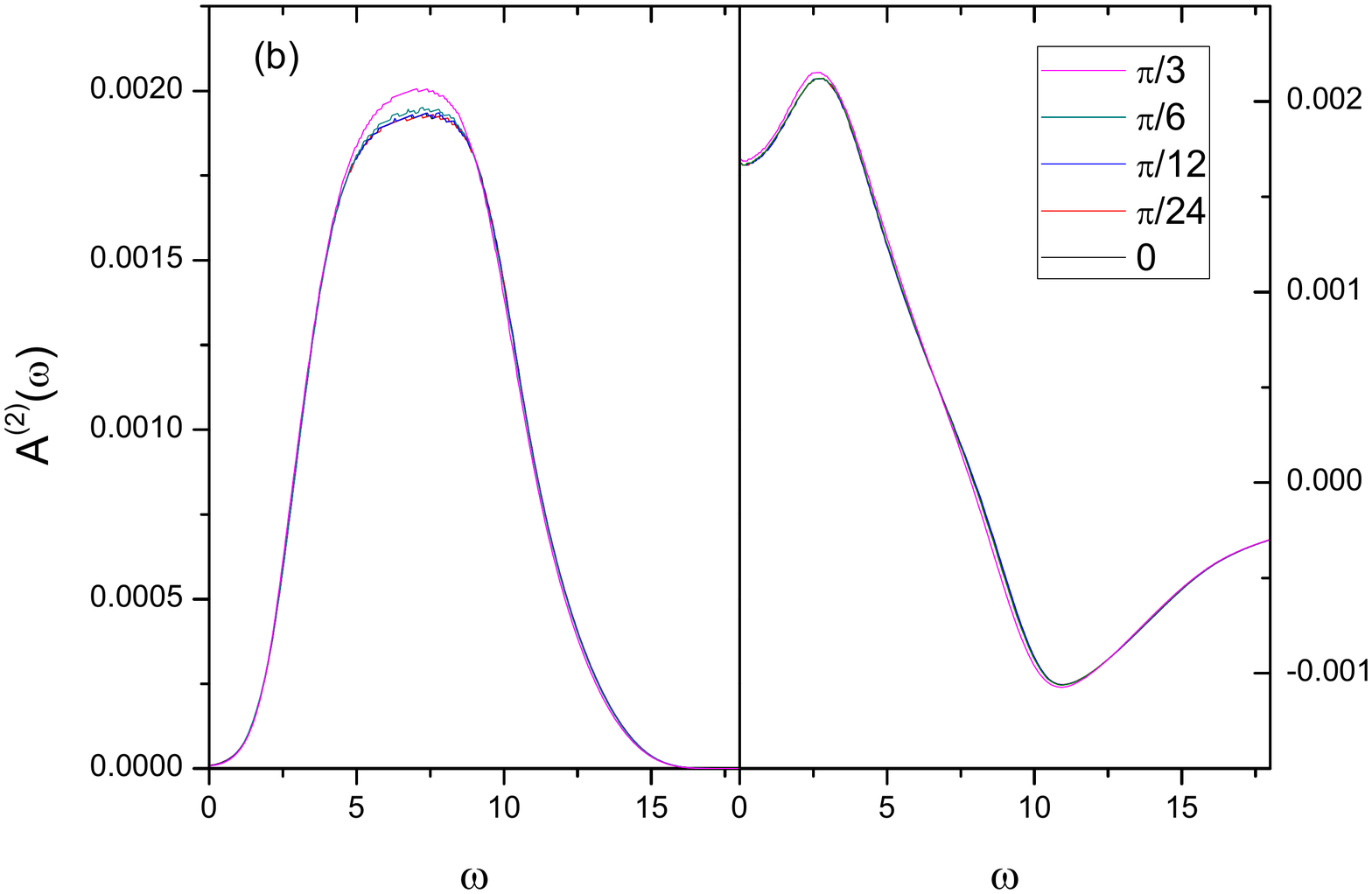}
\end{figure}

\begin{figure}[h!]
\includegraphics[width=16cm,angle=0]{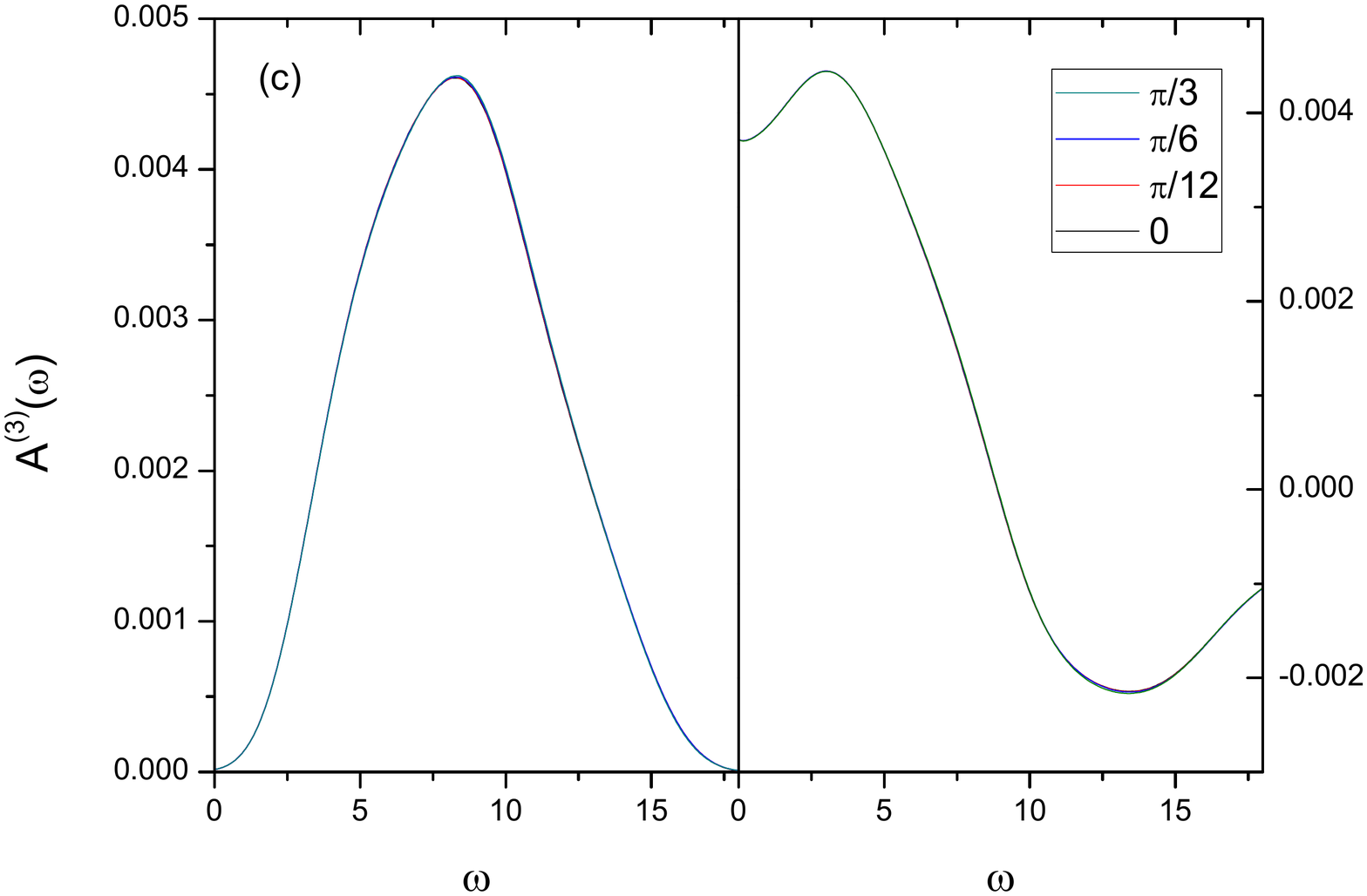}
\caption{The spectral weight(left panel) and the real part(right panel) of the response function of $\hat{C}_{\mathrm{x}}$ on the Fermi sea state from the excitation of (a) one, (b)two and (c)three pairs of particle-hole pairs. Shown here is the result for scalar spin chirality on the up-triangles. We have set the hopping integral of the Fermion between neighboring sites as the unit of energy and adopted the convention $\vec{\mathrm{q}}=q_{x}\vec{\mathrm{G}}_{1}/2+q_{y}\vec{\mathrm{G}}_{2}/2$ for momentum. Here $\vec{\mathrm{G}}_{1,2}$ are the two reciprocal vectors of the triangular lattice.The momentum is chosen at $\mathrm{q}=(q_{x},0)$, with $q_{x}=0,\pi/24,\pi/12,\pi/6$ and $\pi/3$. The calculation of $A^{(1)}(\omega)$ is done in the thermodynamic limit. The calculation of $A^{(2)}(\omega)$ is done on a $48\times48$ lattice. The calculation of $A^{(3)}(\omega)$ is done on a $24\times24$ lattice and $q_{x}=\pi/24$ is inaccessible in this case.}
\end{figure}

\begin{figure}[h!]
\includegraphics[width=16cm,angle=0]{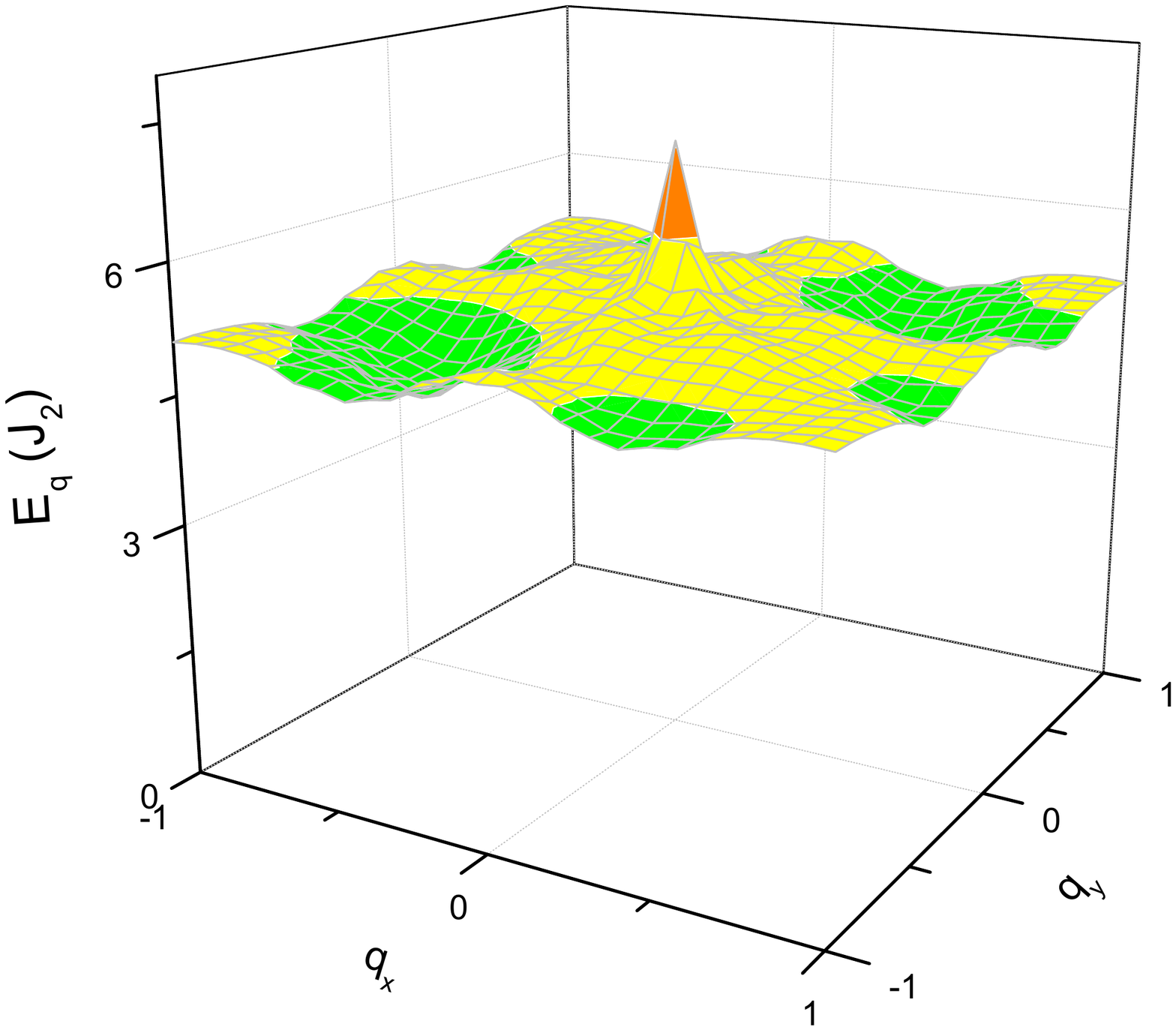}
\caption{The dispersion of the center of gravity of the fluctuation spectrum of the scalar spin chirality in the optical channel. The computation is done on a $24\times 24$ lattice with periodic - antiperiodic boundary condition. We have adopted the convention $\vec{\mathrm{q}}=q_{x}\vec{\mathrm{G}}_{1}/2+q_{y}\vec{\mathrm{G}}_{2}/2$ for momentum, in which $\vec{\mathrm{G}}_{1,2}$ are the two reciprocal vectors of the triangular lattice.}
\end{figure}

\begin{figure}[h!]
\includegraphics[width=16cm,angle=0]{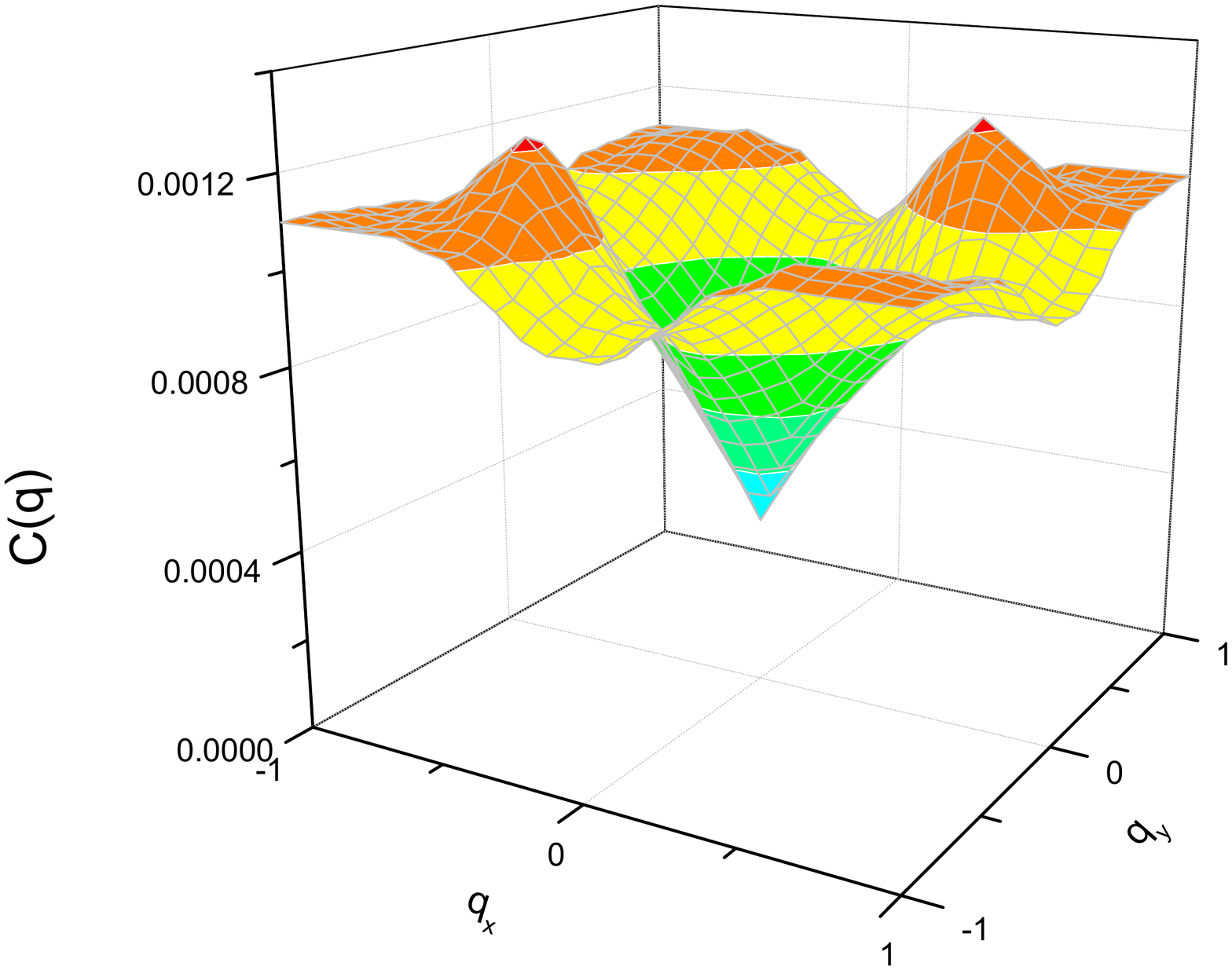}
\caption{The structure factor of the scalar spin chirality in the optical channel. The computation is done on a $24\times 24$ lattice with periodic - antiperiodic boundary condition.}
\end{figure}

\begin{figure}[h!]
\includegraphics[width=16cm,angle=0]{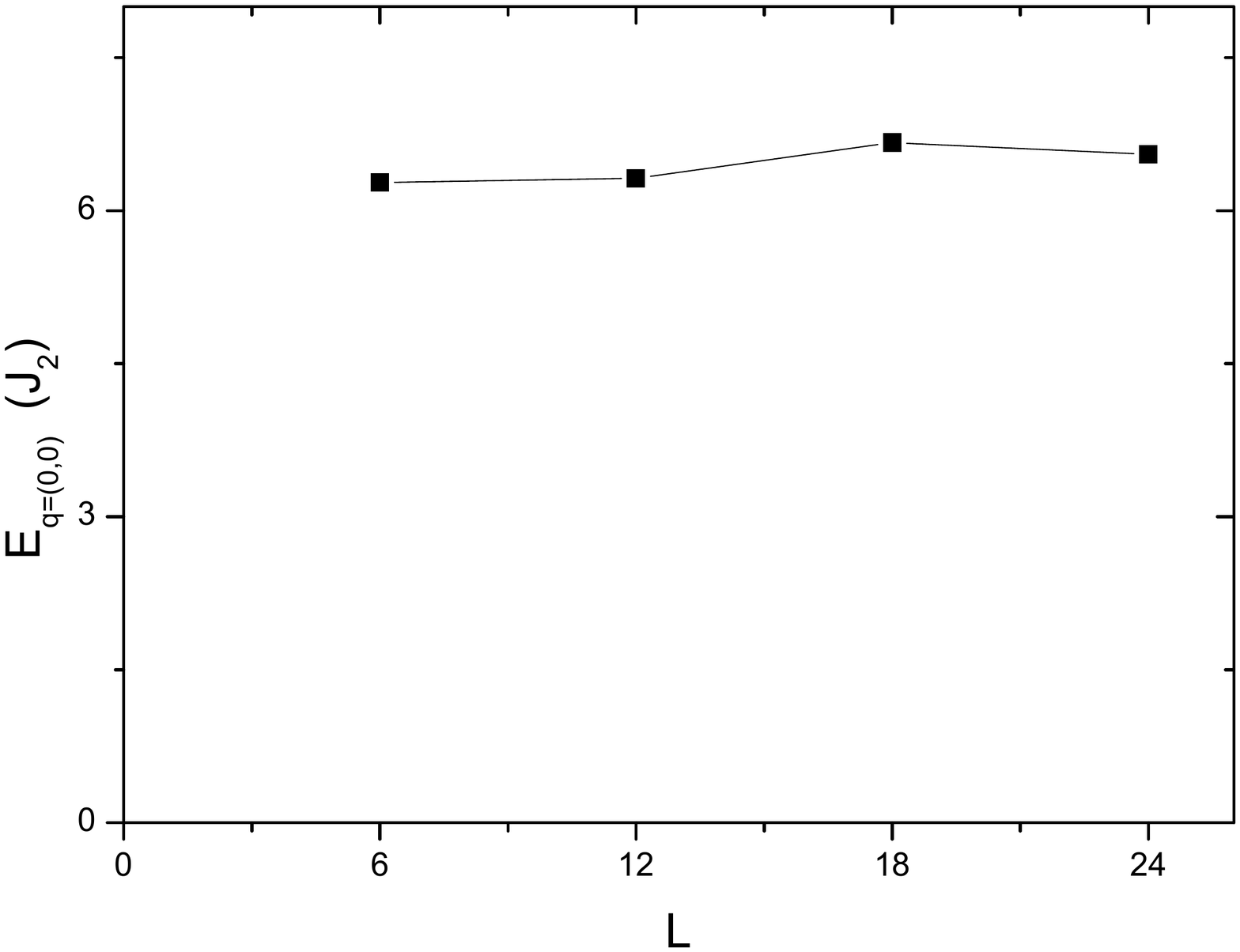}
\caption{The dependence of $E_{\mathrm{q=0}}$ on the lattice size $L$ for the acoustic gauge mode.}
\end{figure}

\begin{figure}[h!]
\includegraphics[width=16cm,angle=0]{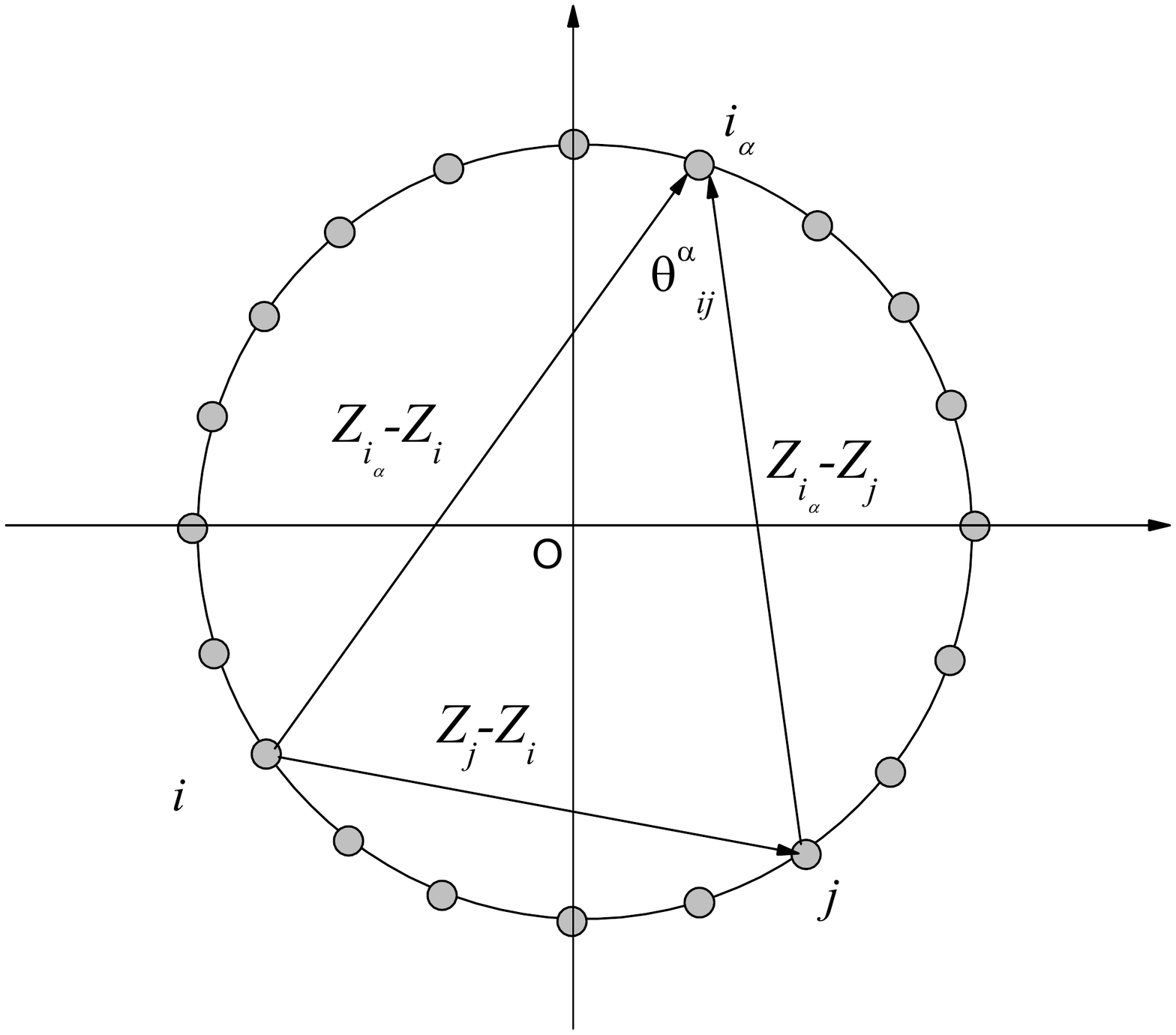}
\caption{The chord coordinate on a ring and the meaning of $\theta^{\alpha}_{i,j}$.}
\end{figure}

\end{document}